\documentclass[lettersize,journal]{IEEEtran}
\usepackage{amsmath,amsfonts}
\usepackage{algorithm}
\usepackage{algorithmic}
\usepackage{array}
\usepackage[caption=false,font=normalsize,labelfont=sf,textfont=sf]{subfig}
\usepackage{amsmath}
\usepackage{textcomp}
\usepackage{stfloats}
\usepackage{url}
\usepackage{verbatim}
\usepackage{graphicx}
\usepackage{cite}
\usepackage{multirow}
\usepackage{hyperref}
\usepackage{tikz}
\usetikzlibrary{positioning}
\usepackage{physics}
\usepackage{tikz}
\usepackage{mathdots}
\usepackage{yhmath}
\usepackage{cancel}
\usepackage{color}
\usepackage{siunitx}
\usepackage{multirow}
\usepackage{amssymb}
\usepackage{gensymb}
\usepackage{tabularx}
\usepackage{extarrows}
\usepackage{booktabs}
\usetikzlibrary{fadings}
\usetikzlibrary{patterns}
\usetikzlibrary{shadows.blur}
\usetikzlibrary{shapes}
\usepackage{diagbox}
\usepackage{bbding}
\usepackage{amsthm}
\usepackage{float}

\theoremstyle{definition}

\newcounter{mynum2}
\newcounter{mynum3}
\newcounter{mynum4}
\newcounter{mynum5}
\begin{document}

\title{Cross-user activity recognition using deep domain adaptation with temporal relation information}

\author{
Xiaozhou Ye
\thanks{Xiaozhou Ye is with the Department of Electrical, Computer and Software Engineering, The University of Auckland, New Zealand (e-mail: xye685@aucklanduni.ac.nz)}
\href{https://orcid.org/0000-0002-9725-1548}{\includegraphics[scale=0.01]{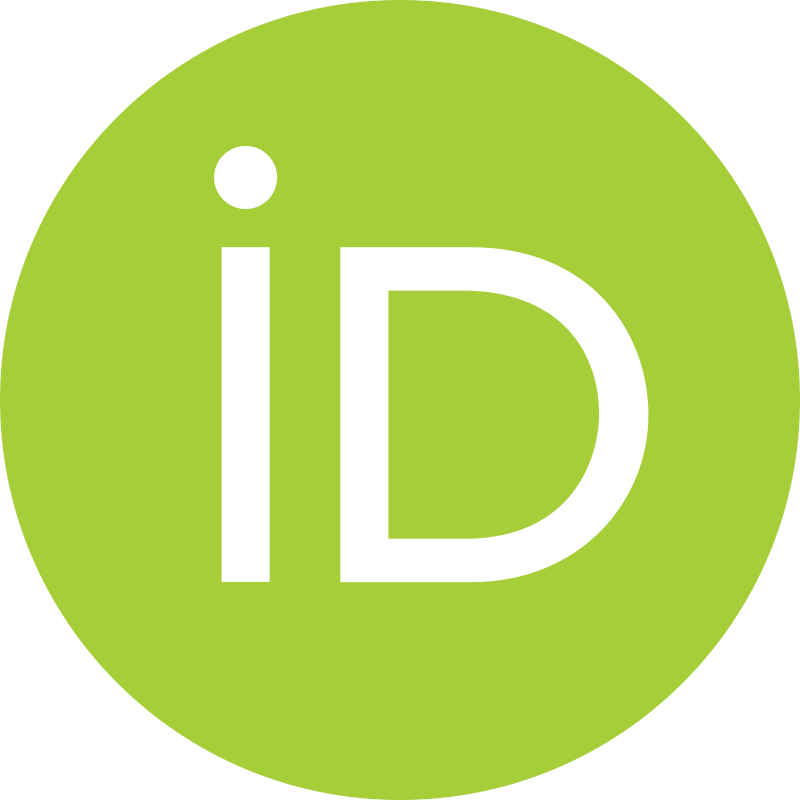}}
\and,
Waleed H. Abdulla
\thanks{Waleed H. Abdulla is with the Department of Electrical, Computer and Software Engineering, The University of Auckland, New Zealand (e-mail: w.abdulla@auckland.ac.nz)}
\href{https://orcid.org/0000-0002-1812-4285}{\includegraphics[scale=0.01]{ORCID_iD.png}}
\and,
Nirmal Nair
\thanks{Nirmal Nair is with the Department of Electrical, Computer and Software Engineering, The University of Auckland, New Zealand (e-mail: n.nair@auckland.ac.nz)}
\href{https://orcid.org/0000-0002-8456-3999}{\includegraphics[scale=0.01]{ORCID_iD.png}}
\and,
Kevin I-Kai Wang
\thanks{Kevin I-Kai Wang is with the Department of Electrical, Computer and Software Engineering, The University of Auckland, New Zealand (e-mail: kevin.wang@auckland.ac.nz)}
\href{https://orcid.org/0000-0001-8450-2558}{\includegraphics[scale=0.01]{ORCID_iD.png}}}

\maketitle

\begin{abstract}
Human Activity Recognition (HAR) is a cornerstone of ubiquitous computing, with promising applications in diverse fields such as health monitoring and ambient assisted living. Despite significant advancements, sensor-based HAR methods often operate under the assumption that training and testing data have identical distributions. However, in many real-world scenarios, particularly in sensor-based HAR, this assumption is invalidated by out-of-distribution ($\displaystyle o.o.d.$) challenges, including differences from heterogeneous sensors, change over time, and individual behavioural variability. This paper centres on the latter, exploring the cross-user HAR problem where behavioural variability across individuals results in differing data distributions. To address this challenge, we introduce the Deep Temporal State Domain Adaptation (DTSDA) model, an innovative approach tailored for time series domain adaptation in cross-user HAR. Contrary to the common assumption of sample independence in existing domain adaptation approaches, DTSDA recognizes and harnesses the inherent temporal relations in the data. Therefore, we introduce 'Temporal State', a concept that defined the different sub-activities within an activity, consistent across different users. We ensure these sub-activities follow a logical time sequence through 'Temporal Consistency' property and propose the 'Pseudo Temporal State Labeling' method to identify the user-invariant temporal relations. Moreover, the design principle of DTSDA integrates adversarial learning for better domain adaptation. Comprehensive evaluations on three HAR datasets demonstrate DTSDA's superior performance in cross-user HAR applications by briding individual behavioral variability using temporal relations across sub-activities.

\end{abstract}


\section{Introduction}

Human activity recognition (HAR) is a pivotal research field in human-computer interaction, ubiquitous computing, and the Internet of Things, with the objective of accurately classifying human activities based on sensor data and context information \cite{hussain2020review}. HAR finds extensive applications in areas such as medical treatment, ambient assisted living, fitness, sports, rehabilitation, security surveillance, health monitoring, and home automation \cite{yadav2021review}. Despite the research effort, several challenges still remain in achieving an effective HAR model. Most of the prevailing HAR methods that process time series sensor data operate under the assumption that training and testing data (in the target application) are drawn from the same distribution, ensuring the model's generalizability. This presumes that the training and testing data come from the same domain with identical feature space and data distribution characteristics. However, in many practical scenarios, this assumption does not hold due to data heterogeneity or $\displaystyle o.o.d. $ data \cite{chen2021deep}. Consequently, a model trained on the source domain may perform poorly on the target domain, leading us to a more practical challenge: the data distributions of the training and testing sets may differ. 

Sensor-based HAR experiences several $\displaystyle o.o.d. $ challenges across various categories. Different sensors, stemming from variations in manufacturers, types, and purposes, can lead to heterogeneous data formats and distributions \cite{xing2018enabling}. Additionally, the phenomenon of concept drift signifies that data patterns can evolve over time, adding a layer of complexity to HAR \cite{lu2018learning}. Individual behavior also introduces variability, as people inherently display diverse activity patterns \cite{saeedi2018personalized}. Further, the placement of sensors, whether worn on the body \cite{rokni2018autonomous} or positioned within smart home environments \cite{sukhija2019supervised}, can alter the data due to differences in positions or setups. Our current research is particularly focusing on addressing the $\displaystyle o.o.d. $ challenge in sensor-based HAR that arise due to the behavioural variability between different individuals, i.e., the cross-user HAR problem.

Transfer learning domain adaptation \cite{farahani2021brief} presents a promising direction to handle cross-user HAR. It aims to identify shared knowledge and diminishes the data distribution differences between the source and target domains. For instance, Maximum Mean Discrepancy (MMD) \cite{yan2017mind} aligns the source and target domains by minimizing the distance between their means in a Reproducing Kernel Hilbert Space. Correlation Alignment (CORAL) \cite{sun2016return} adapts source domain data to a target domain by aligning the second-order statistics (covariances) of their feature distributions. While traditional methods have their merits, deep learning has ushered in a slew of deep domain adaptation techniques that discovers the intricate relations hidden in data. Transferable Semantic Augmentation \cite{li2021transferable} creates a distribution based on the differences in deep feature means between domains and the variability within the target domain's features for modifying the source deep features to align more closely with the target's semantics. Simultaneous Semantic Alignment Network \cite{li2020simultaneous} works by closely aligning the centroids of data distributions from both the source and target domains using a non-linear deep mapping. 

Most of the existing domain adaptation research as mentioned above, focuses on static data, such as in computer vision, where samples are assumed to be $\displaystyle i.i.d. $ within its corresponding domain. Deep Domain Adaptation model for Time Series data \cite{wilson2020multi} focuses on sensor-based HAR domain adaptation. However, the samples $\displaystyle i.i.d. $ assumption within its corresponding domains still applied to the domain adaption methods. In fact, time series data often possess inherent temporal relation, suggesting that each sample is not independent. It's evident that while there has been significant progress in domain adaptation techniques, there remains a gap in fully leveraging the temporal relation knowledge in time series data. Such insights could be pivotal for improved time series domain adaptation for cross-user HAR task.

In this paper, we introduce the Deep Temporal State Domain Adaptation (DTSDA) method to target specifically time series domain adaptation for cross-user HAR. This method accounts for temporal relations over time, contrasting with previous research that assumed independence among samples. The capture and utilization of temporal relations rely on two aspects. First, a new concept called temporal state is proposed for describing different phases of an activity considering the sub-activities embedded in the human motion. For example, the walking activity includes three sub-activities: raising the leg, thrusting forward and feet to the ground. Such temporal state is common knowledge across users, which means different users’ time series activity data should follow the same temporal states in their corresponding activities. These temporal states can be used as a bridge for achieving more accurate cross-user HAR.  

Second, the temporal consistency property is introduced for better temporal relation learning and temporal state acquisition. Temporal consistency in HAR refers to the sub-activities that transit sequentially from one to the next over time. This is important because human activities typically follow logical sequences, and inconsistencies in the sub-activities can lead to misclassification and have adverse impact on domain adaptation. For example, in making tea, the tea bag should be removed only after steeping, not before; reversing this sequence would violate the logical sequence of the activity. Considering the temporal consistency property, a novel self-supervised learning annotation method called Pseudo Temporal State Labeling is proposed.

\begin{figure}[h!]
\centering
\includegraphics[width=\columnwidth]{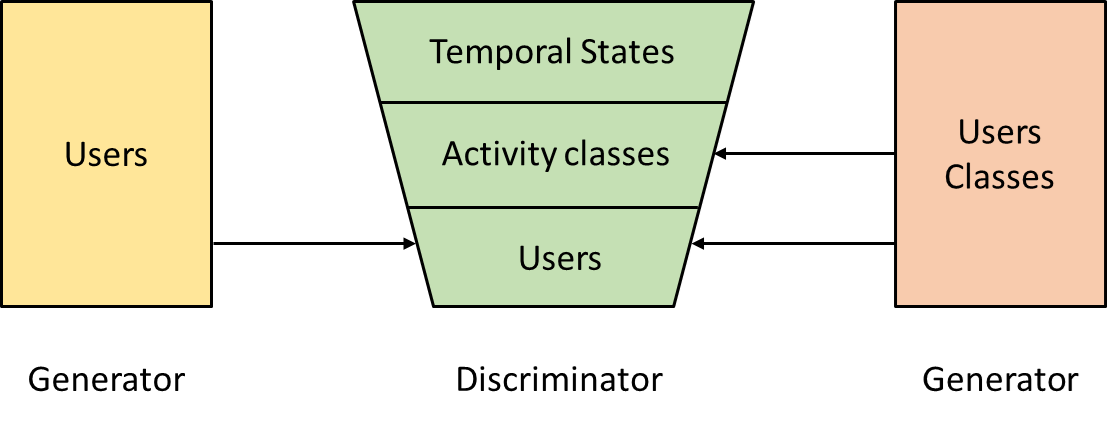}
\caption{The design principle overview of DTSDA.\label{GAN_framework_explain}}
\end{figure}

For the extraction of user-invariant temporal states, the DTSDA method integrates the concept of adversarial learning as shown in Figure~\ref{GAN_framework_explain}. There are two main components of DTSDA that function similarly to the generator and discriminator in Generative Adversarial Network (GAN) domain adaptation \cite{zhang2021universal}. The discriminator works hard to identify temporal states, users, and activity classes, while the generator on the right aims to obscure the data distributions of users and activity classes. Through the adversarial training process, it becomes possible to learn temporal states that are invariant to both users and activity classes. Subsequently, using this user-invariant temporal state information, the third component of DTSDA acts as another generator as the one on the left. This generator works in conjunction with the discriminator to blur the data distributions of users, aiding in the learning of a user-invariant activity classifier. This process ultimately contributes to improved cross-user Human Activity Recognition (HAR) tasks.

In summary, the contributions of the work:

\begin{enumerate}
\item A novel method called DTSDA, specifically tailored for time series deep domain adaptation in cross-user HAR, is proposed that integrates adversarial learning and leverages the temporal relation knowledge present in time series data. 
\item A new concept called 'Temporal State' is proposed to capture the various phases of an activity, taking into account the common consecutive sub-activities.
\item A novel self-supervised learning annotation technique called 'Pseudo Temporal State Labeling' is proposed by making use of the temporal consistency property in time series.
\end{enumerate}

The subsequent sections of this paper are organized as follows. Section \Roman{mynum2} presents the related work, studying the approaches of cross-user HAR and introducing the fundamental principles of transfer learning and domain adaptation. Section \Roman{mynum3} provides a comprehensive explanation of the proposed method, elaborating on the problem formulation and the design details of the DTSDA, with a spotlight on temporal relation representations and the deep neural network design for data distribution alignment. Section \Roman{mynum4} pivots towards delineating the experimental setup and providing a comparative analysis between our approach and existing methods. The effect of temporal relation knowledge is explored. Concluding our discourse, Section \Roman{mynum5} draws inferences from our explorations and sets the direction for potential future research.

\section{Related work}

\subsection{Cross-user human activity recognition}

HAR is an essential component of ubiquitous computing due to its significant role in our everyday life. Its purpose is to analyze and identify human activities by leveraging higher-level knowledge gathered from a range of sensor data and contextual information. Numerous sensor types are employed for activity recognition across various situations. From a sensor modality standpoint, HAR can be grouped into five categories: HAR based on smartphones/wearables, ambient sensors, device-free sensors, vision-based sensors, and other modalities, as reported in several studies \cite{chen2021deep}\cite{lentzas2020non}\cite{zhang2021privacy}\cite{roche2021multimodal}\cite{pan2020hierarchical}. In this study, we primarily focus on HAR using wearable sensors.

From the perspective of machine learning research, sensor-based HAR is viewed as a time series classification problem \cite{yadav2021review}. A plethora of classification models, such as ensemble learning \cite{sekiguchi2020ensemble}, Support Vector Machine \cite{shuvo2020hybrid}, and Hidden Markov model (HMM) \cite{liu2021motion} have been proposed to tackle the HAR challenge. Concurrently, with the advancement of research, deep learning has shown superior results across numerous tasks. Deep learning-based HAR methods \cite{duan2023multi} can learn high-level features and autonomously extract characteristics from extensive data sets. However, these methods largely rely on the assumption that the training and testing data are derived from the same distribution, thus ensuring the model's generalizability, i.e., all data is independently and identically distributed ($\displaystyle i.i.d.$) \cite{chen2021deep}. This assumption doesn't hold in several real-world applications, where the training and testing datasets are often out-of-distribution ($\displaystyle o.o.d.$) \cite{lu2022out}. This study focuses on the sensor-based HAR problem in an $\displaystyle o.o.d.$ context.

There are multiple variants of the sensor-based HAR $\displaystyle o.o.d. $ problem: firstly, data can originate from diverse sensors, with different types, platforms, manufacturers, and modalities potentially resulting in distinct data formats and distributions \cite{xing2018enabling}. Secondly, data patterns may change over time, a phenomenon also known as concept drift \cite{lu2018learning}. An individual's walking pattern, for instance, could vary due to health conditions. Thirdly, there can be significant behaviour differences between different individuals \cite{saeedi2018personalized}, for example, different walking speeds. Lastly, the location of physical sensors, whether on different parts of the body \cite{rokni2018autonomous} or in varied configurations within smart homes \cite{sukhija2019supervised}, can lead to distinct data distributions. This study primarily addresses the sensor-based HAR $\displaystyle o.o.d. $ problem concerning behavioural differences across individuals.

\subsection{Transfer learning and domain adaptation}

Transfer learning \cite{zhuang2020comprehensive} is a technique that allows a model to be trained in one or several source domains and then applied to one or more related target domains, which typically have no or limited labels. A crucial task during this process is addressing the $\displaystyle o.o.d. $ problem to minimize the distribution disparity between the source and target domains. Domain adaptation \cite{wilson2020survey}, a specific type of transfer learning, is also aimed at addressing the $\displaystyle o.o.d. $ issue between source and target domains but under the added constraint that the tasks in these domains are the same. The primary focus of this paper is the domain adaptation problem, characterized by the presence of labelled data from the source domain and unlabeled data from the target domain.

Domain adaptation has undergone significant evolution, with feature-based domain adaptation emerging as a predominant approach \cite{farahani2021brief}. At its core, this method emphasizes the importance of feature representations. The central premise is to transfer knowledge encapsulated in these features from a source domain to a target domain. The goal is to align the feature distributions of the source and target domains so that a model trained on the source domain can perform well on the target domain. This is often done by finding a shared feature space where the distributions of the source and target domains are similar or by transforming one domain so its feature distribution matches the other.

The Adaptive Component Embedding (ACE) \cite{jing2020adaptive} aims to align the subspaces of the source and target domains. This subspace is where the first-order and the second-order statistics of the domains align, and the geometric properties of the original data are preserved. After alignment and embedding, the features are classified by optimizing a structural risk functional in the Reproducing Kernel Hilbert Space. The Dual-Representation Autoencoder (DRAE) \cite{yang2021dual} is designed to learn two types of representations: global and local.  Global representations help in aligning overall distributions between source and target domains, while local representations preserve class-specific information. By combining these representations, DRAE ensures both domain invariance and class discriminativeness, aiming to prevent performance degradation due to loss of class-specific information. Optimal Transport for Domain Adaptation (OTDA) \cite{flamary2016optimal} introduces an approach to address the domain adaptation challenge through the Optimal Transport theory. It identifies a transformation that bridges the source and target domains by ensuring the transportation cost, based on a designated measure, is kept to a minimum. Substructural Optimal Transport (SOT) \cite{lu2021cross} also employs Optimal Transport theory to tackle the domain adaptation problem, which considers not only coupling between two probability density functions but also substructure-level mapping. 

Deep domain adaptation, given its ability to combine deep learning's automated feature learning with domain adaptation's cross-domain knowledge transfer, is a promising avenue for research. Multiple-view Adversarial Learning Network (MAN)  \cite{gao2021novel} utilizes adversarial training to force the discriminator to be unable to differentiate between domains. The ultimate goal is to promote the development of high-quality domain-invariant features. Deep Joint Distribution Alignment (DJDA) \cite{qin2022deep} is a deep domain adaptation technique that is designed to minimize the discrepancies between the source and target domains. It works on aligning both marginal and conditional distributions of these two domains. 

These feature-based domain adaptation techniques largely focus on static data like images \cite{yang2021advancing}\cite{li2020simultaneous} and often unify time series data into the same domain adaptation framework. However, these existing approaches may not be fully effective with sensor-based HAR data since they tend to overlook the temporal relation knowledge intrinsic to time series data during the alignment of data distribution. Temporal relation is a critical attribute of time series data, and solely relying on temporal invariant knowledge for time series domain adaptation might lead to suboptimal model performance. Moreover, temporal relation knowledge could potentially help identify common knowledge between users, further improving sensor-based HAR domain adaptation.

\section{Deep Temporal State Domain Adaptation}

\subsection{Problem formulation}
In a cross-user HAR problem, a labelled source user $\displaystyle S^{Source} =\left\{\left( x_{i}^{Source} ,\ y_{i}^{Source}\right)\right\}_{i=1}^{n^{Source}} $ drawn from a joint probability distribution $\displaystyle P^{Source}(X, y) $ and a target user $\displaystyle S^{Target} =\left\{\left( x_{i}^{Target} ,\ y_{i}^{Target}\right)\right\}_{i=1}^{n^{Target}} $ drawn from a joint probability distribution $\displaystyle P^{Target}(X, y) $, where $\displaystyle {n^{Source}} $ and $\displaystyle {n^{Target}} $ are the number of source and target samples respectively.  $\displaystyle S^{Source}$ and $\displaystyle S^{Target}$ have the same feature spaces (i.e. the set of features that describe the data from sensor readings) and label spaces (i.e. the set of activity classes). The source and target users have different distributions, i.e., $\displaystyle P^{Source}(X, y) \neq P^{Target}(X, y) $, which means that even for the same activity, the sensor readings look different between the two users. 
Given source user data $\displaystyle \left\{\left( x_{i}^{Source} ,\ y_{i}^{Source}\right)\right\}_{i=1}^{n^{Source}} $ and target user data $\displaystyle \left\{\left( x_{i}^{Target} \right)\right\}_{i=1}^{n^{Target}} $, the goal is to obtain the labels for the target user.

\subsection{DTSDA method overview}
\begin{figure}[h!]
\centering
\includegraphics[width=\columnwidth]{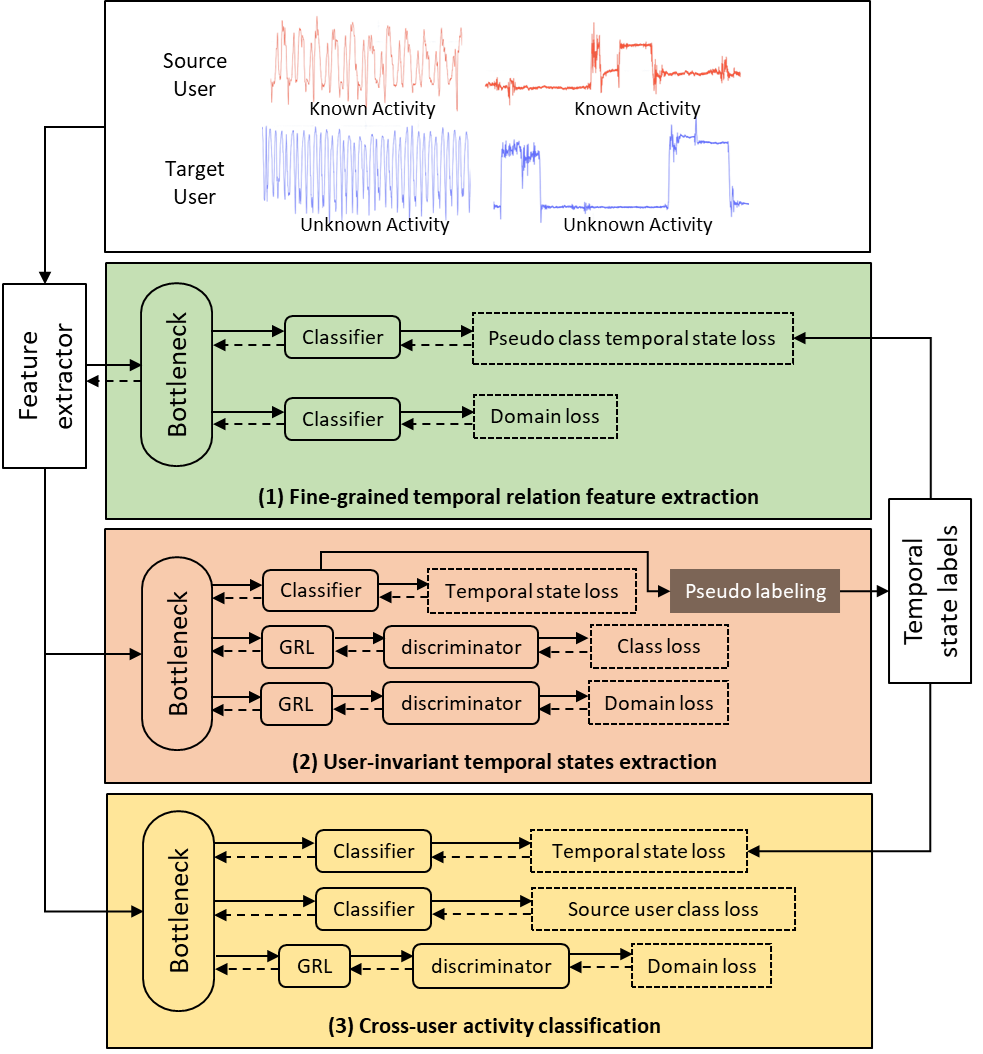}
\caption{The method overview of DTSDA.\label{framework_DGTSDA}}
\end{figure}

In this paper, we propose the DTSDA method to address the cross-user HAR problem. Due to the nature of the physical movement, individual motions have dependencies on earlier motions. We aim to explore the use of temporal relations embedded in human activity time series data. Such temporal relation knowledge may share commonality across users and can be used to enhance the performance and robustness of the developed domain adaptation HAR model. To this end, a deep domain adaptation method with adversarial learning is designed to utilize temporal relation knowledge and perform cross-user HAR. More specifically, the proposed DTSDA consists of the following three iterative components in Figure~\ref{framework_DGTSDA}. These three components are interrelated and collectively address the challenge of cross-user HAR.

\begin{enumerate}
  \item Fine-grained temporal relation feature extraction: The purpose of this component is to extract features that can correctly distinguish the temporal states, activity classes and users (i.e. domains). Its function is similar to that of a discriminator in GAN domain adaptation for cooperatively usage with other two components (act like generator in GAN domain adaptation) to realize adversarial learning for learning user-invariant temporal state extractions and cross-user activity classification. Therefore, the learned feature extractor is used for the subsequent two components.
  \item User-invariant temporal states extraction: This component aims to extract temporal states that are consistent across users, essentially finding patterns in time-series data that aren't affected by individual user variations. It is similar to the generator in GAN domain adaptation that confuses the data distributions of activity classes and users to achieve the goal of user-invariant temporal states extraction in a feature distribution adjustment way cooperated with the first components. In addition, the temporal consistency property is used for better temporal state extraction since human activities follow logical sequences, regardless of the users. By introducing the temporal consistency property, DTSDA can focus on user-agnostic features, which are crucial for domain adaptation. Pseudo Temporal State Labeling method is proposed and applied here as inspired by the idea of temporal consistency property. The learned temporal state labels are used for the other two components.
 
  \item Cross-user activity classification: This component aims to leverage the insights from the first two components to build a classifier that can be generalized acorss users. It acts like the generator in GAN domain adaptation that confuses the data distributions of users to achieve cross-user feature distribution adjustment goal cooperated with the first components. In the same time, the user-invariant temporal relation information learned from the second component is introduced to guide the feature distribution adjustment for better data distribution alignment between source and target users. Then, a robust classifier can be built that uses source user classification labels. This classifier, thus, can effectively label the target user's data, benefiting from the temporal relation insights extracted.
\end{enumerate}

\begin{figure*}[h!]
\centering
\includegraphics[width=0.9\textwidth]{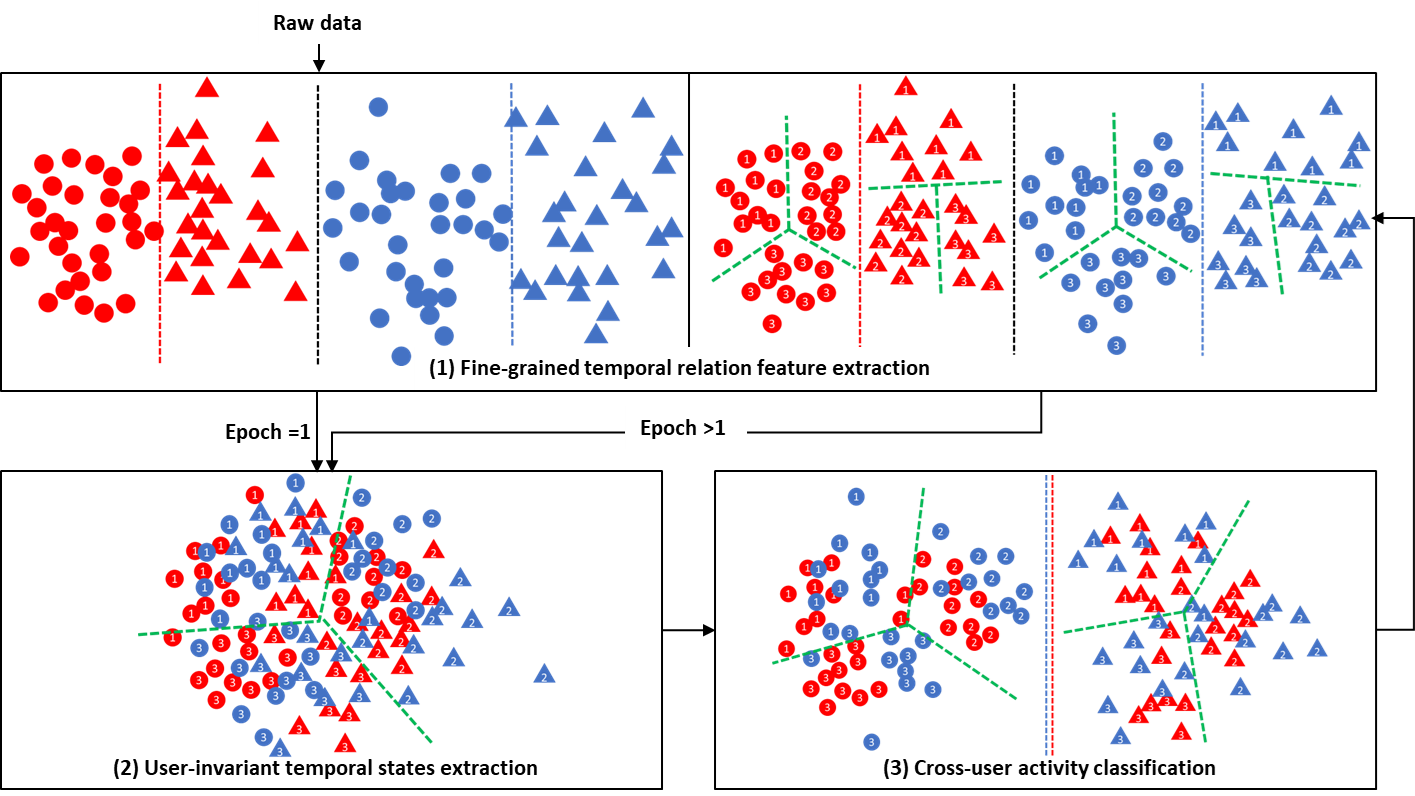}
\caption{The data distribution adjustment overview of DTSDA.\label{DTSDA_schematic_diagram}}
\end{figure*}

In essence, the design of DTSDA aims to bridge the gap between source and target users by leveraging temporal relation information. The three components are sequential and complementary. Component (1) identifies the detailed differences across users, activities and temporal states. Component (2) cooperates with the first component by focusing on similarities rather than differences of users and activities, emphasizing temporal relations that remain consistent across users. Moreover, the learned temporal state labels in component (2) are fed back to component (1) for guiding the fine-grained temporal relation feature extraction in the next iteration. Component (3) then uses the insights from both the differences from component (1) and the similarities from component (2) to build a robust and generalized classifier. The overall method can be treated as a type of adversarial learning that plays the differences and similarities game, and can further enhance the alignment of source and target domain distributions by leveraging temporal relation information. In the following three subsections, the detailed design principle and implementation of the three components are explained. 
 
\subsection{Fine-grained temporal relation feature extraction}

This component aims to design a feature extractor that can learn fine-grained data distributions. The ideal features should help distinguish the different users, activity classes, and temporal states. In this way, the knowledge contained in the data can be fully utilized, which forms the basis for the following components. The learned feature extractor offers the "lens" through which the other two components will view the data. Its findings are instrumental in guiding the extraction of user-invariant temporal states and formulating a classifier that can generalize across users.

Figure~\ref{DTSDA_schematic_diagram} shows example data distributions corresponding to each individual component. This illustration aims to provide an intuitive understanding of the data distribution adjustment and alignment process. Figure~\ref{DTSDA_schematic_diagram} (1) has two parts: the left is the initial setting at the first epoch, and the right represents the subsequent epochs during model training. The red colour is associated with the source user, while the blue is associated with the target user. The different shapes, such as circle and triangle, represents different activity classes. There are two classifiers, one of them called domain classifier as shown in Figure~\ref{framework_DGTSDA} (1), that focuses on identifying the source and target users, as the black dashed line in Figure~\ref{DTSDA_schematic_diagram} (1). The labels for this classifier are available for binary classification, distinguishing between the source and target users. The other classifier, called pseudo class temporal state classifier in Figure~\ref{framework_DGTSDA} (1), is used to classify the different classes and the corresponding temporal states. 
For this classifier, only the source user labels are available, while the target user labels are unavailable. Although the target domain label is unknown, it is common to assume that the association between samples and classes is known \cite{uslu2022segmentation}\cite{hallac2017toeplitz}\cite{hallac2019greedy}. For example, we may know a particular sample is associated with class A, but we have no knowledge of what class A is (i.e. walking, running, etc.).

\begin{figure}[h!]
\centering
\includegraphics[width=\columnwidth]{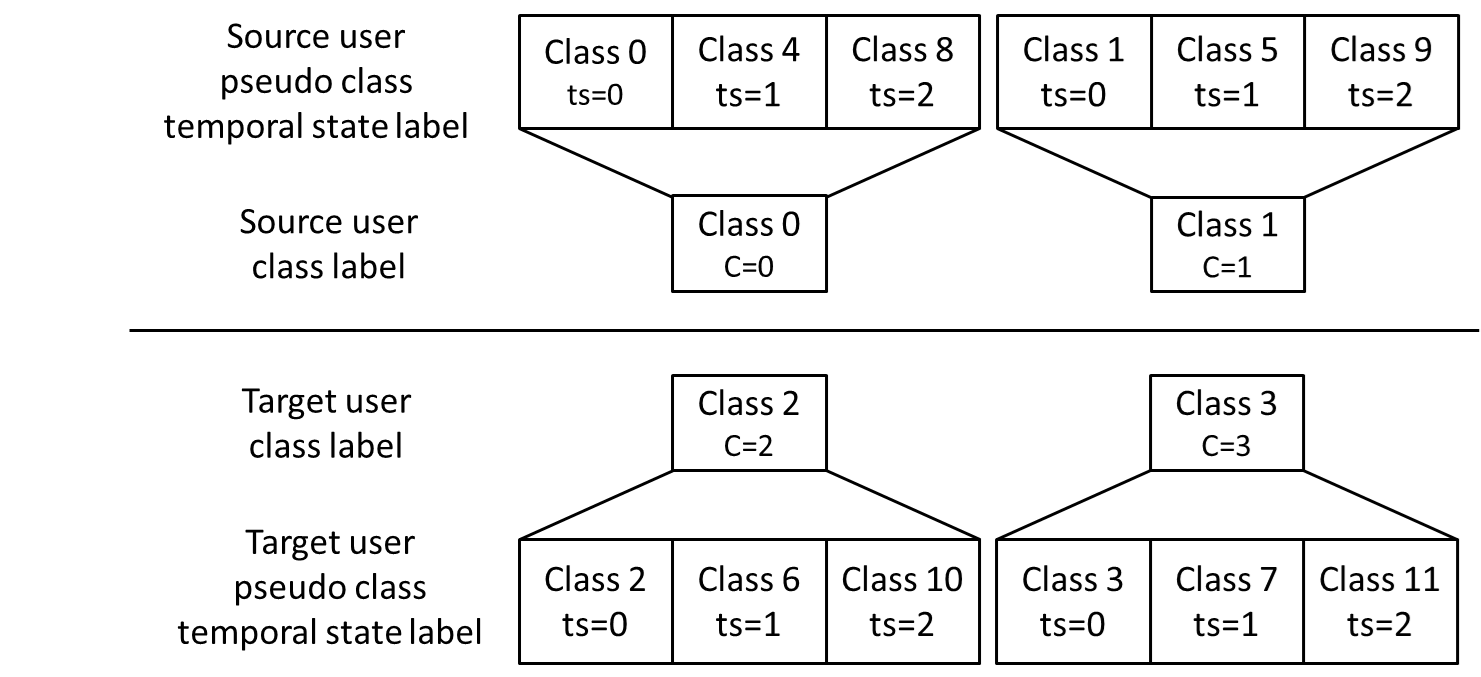}
\caption{An example of pseudo class temporal state label .\label{pseudo_class_temporal_state_label}}
\end{figure}

Therefore, we can set a number as a label for each of the classes in the target user, but the number is set different from the label numbers in the source user. For convenience, we arrange the number $\displaystyle [{C}, {2\times C-1}]$ for target user labels, where $\displaystyle {C}$ is the number of classes in each user. Now, we have a total of $\displaystyle {2\times C}$ class labels for the source and target users, the class labels of all the samples is $\displaystyle c\in \{0,1,2,...,2\times C-1\}$. During training, the pseudo class temporal state label is assigned as $\displaystyle \hat{y} =ts\times 2\times C+c$, where $\displaystyle ts$ is the temporal state labels of all the samples $\displaystyle ts\in \{0,1,...,T-1\},$ where $\displaystyle T$ is a pre-defined hyper-parameter that represents the number of temporal states. Figure~\ref{pseudo_class_temporal_state_label} is an example of the implementation of pseudo class temporal state label given $\displaystyle {C}=2$ and $\displaystyle T=3$.

In the first epoch, the temporal states have not yet been learned in component (2) and we simply set the label of temporal state for all samples to be 0, i.e., $\displaystyle ts= \{0,0,...,0\}$ as shown in the initial setting illustrated in the left part of Figure~\ref{DTSDA_schematic_diagram} (1), the red dashed line is the decision boundary of activity classes of source user while the blue dashed line is the decision boundary of target user. Once the temporal state labels $\displaystyle ts$ are available from the component (2) in the subsequent epochs, the temporal state information is used for feature extraction as illustrated on the right hand side of Figure~\ref{DTSDA_schematic_diagram} (1). The green dashed lines are boundaries that separate different temporal states within each activity class. The number in each shape indicates the corresponding temporal state information hidden in the time series data. For example, the numbers one, two and three in triangles may correspond to raising the leg, thrusting forward and feet to the ground of walking activity. 

\begin{figure}[H]
\centering
\includegraphics[width=\columnwidth]{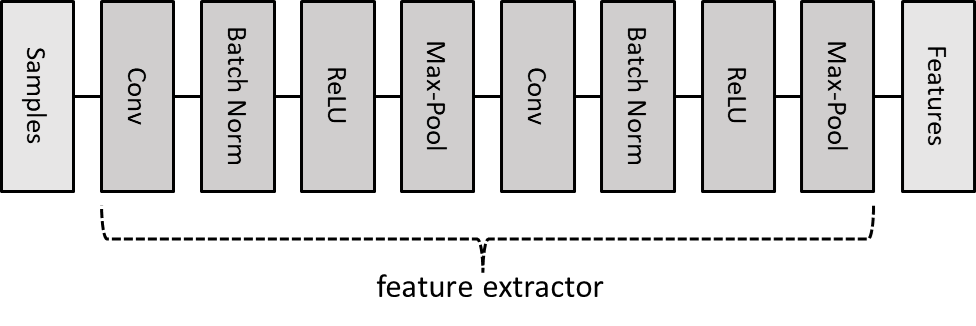}
\caption{The network architecture of feature extractor module.\label{feature_extraction}}
\end{figure}

The network architecture design of the feature extraction module is shown in Figure~\ref{feature_extraction}. The feature extractor comprises two main convolutional layers. Each layer consists of a 1D convolution layer for temporal feature extraction due to the time series samples, followed by batch normalization for reducing the internal covariate shift, a ReLU activation function for non-linearity mapping, and a max-pooling operation for downsampling. After processing the input samples, the learned features are outputted.

\begin{figure}[H]
\centering
\includegraphics[width=0.6\columnwidth]{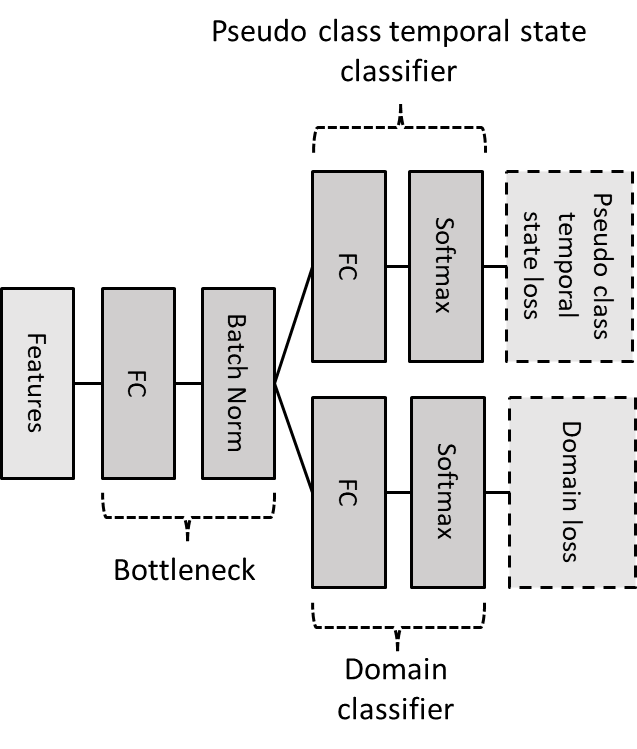}
\caption{The network architecture of fine-grained temporal relation feature extraction.\label{Fine_grained_distribution_distinctions}}
\end{figure}

The network architecture of the fine-grained temporal relation feature extraction component is shown in Figure~\ref{Fine_grained_distribution_distinctions}. The network architecture begins by taking input features and passing them through a bottleneck layer for potentially reducing user-specific characteristics by learning a compact representation of features. Specifically, within this layer, the data undergoes a linear transformation to condense its size. Following this transformation, batch normalization is applied as a regularizer. After the bottleneck processing, the condensed features are directed towards two distinct classifiers. The first one is the pseudo class temporal state classifier, which utilizes a linear layer to output multiple classes. Simultaneously, the same features are fed into a second classifier, which serves as a domain classifier, predicting from two classes, i.e. source and target users. This classifier also uses a simple linear transformation layer followed by a softmax layer.

Here, we explain the design of loss functions. $\displaystyle h_{f} $ is the feature extractor for the global method. $\displaystyle h_{bf} $, $\displaystyle h_{cf}^{A} $, and $\displaystyle h_{cf}^{B} $ are the bottleneck, pseudo class temporal state classifier, and domain classifier of fine-grained temporal relation feature extraction component.

Pseudo Class Temporal State Loss:
\begin{equation}
L_{pctsf} = l_{CE}\left( h_{cf}^{A}( h_{bf}( h_{f}( x))),\hat{y}\right)
\end{equation}

Domain Loss:
\begin{equation}
L_{df} = l_{CE}\left( h_{cf}^{B}( h_{bf}( h_{f}( x))) ,d\right)
\end{equation}

The Total Loss:
\begin{equation}
 L_{f} = L_{pctsf} + L_{df}
\end{equation}

Where $\displaystyle l_{CE}$ is the cross-entropy loss function. $\displaystyle d $ is the domain labels, $\displaystyle d\in \{0,1\}$, 0 and 1 represent source and target users separately.

\subsection{User-invariant temporal states extraction}

\subsubsection{The overall implementation process}

This component is about discerning temporal relations that remain consistent across different users based on the idea that sub-activity sequences are common across different individuals yet manifest in unique, user-specific ways. The user-invariant temporal states extraction component can find the different sub-distributions and the corresponding temporal relations in each class to describe the common sub-activity sequence of the activities across users. 

The interaction between component (1) and this component in adversarial learning way, analogy to discriminator and generator in GAN domain adaptation, discriminator learns to identify different data distributions while generator aims to make the data distributions indistinguishable. As the feature extraction process in component (1) becomes more discriminative for activity classes and users, and this component highlights commonalities across users and activity classes, essentially making the data distributions more similar. This push-and-pull dynamic is central to adversarial learning and is what drives the improvement in feature representation of component (1) and data distribution adjustment of this component.

The commonality learned in this component establishes a baseline for understanding human activities in a way that is agnostic to individual variations. Therefore, the learned user-invariant temporal relation knowledge can be used to guide the other two steps. When component (3) is building a classifier, having this baseline ensures that the classifier doesn't get too tailored to the nuances of a specific user, which enhances its generalizability.

For this component itself, adversarial learning \cite{ganin2016domain} is also applied to ensure the temporal states learned are robust, generalizable, and not overfitted to specific users. The Gradient Reversal Layer (GRL) \cite{ganin2016domain} is an ingenious adversarial learning technique that inverts the gradients being backpropagated to encourage the component to adjust the feature distribution that is not only distinctive for the temporal states but also invariant across activity classes and domains. As in Figure~\ref{framework_DGTSDA} (2) and Figure~\ref{DTSDA_schematic_diagram} (2), the temporal state loss aims to correctly categorize the temporal states. Simultaneously, the class loss and domain loss strive to confuse the class and domain discriminators. This confusion forces the feature mapping learning process to be user-invariant and class-invariant as shown in the example distributions in Figure~\ref{DTSDA_schematic_diagram} (2).

\begin{figure}[H]
\centering
\includegraphics[width=\columnwidth]{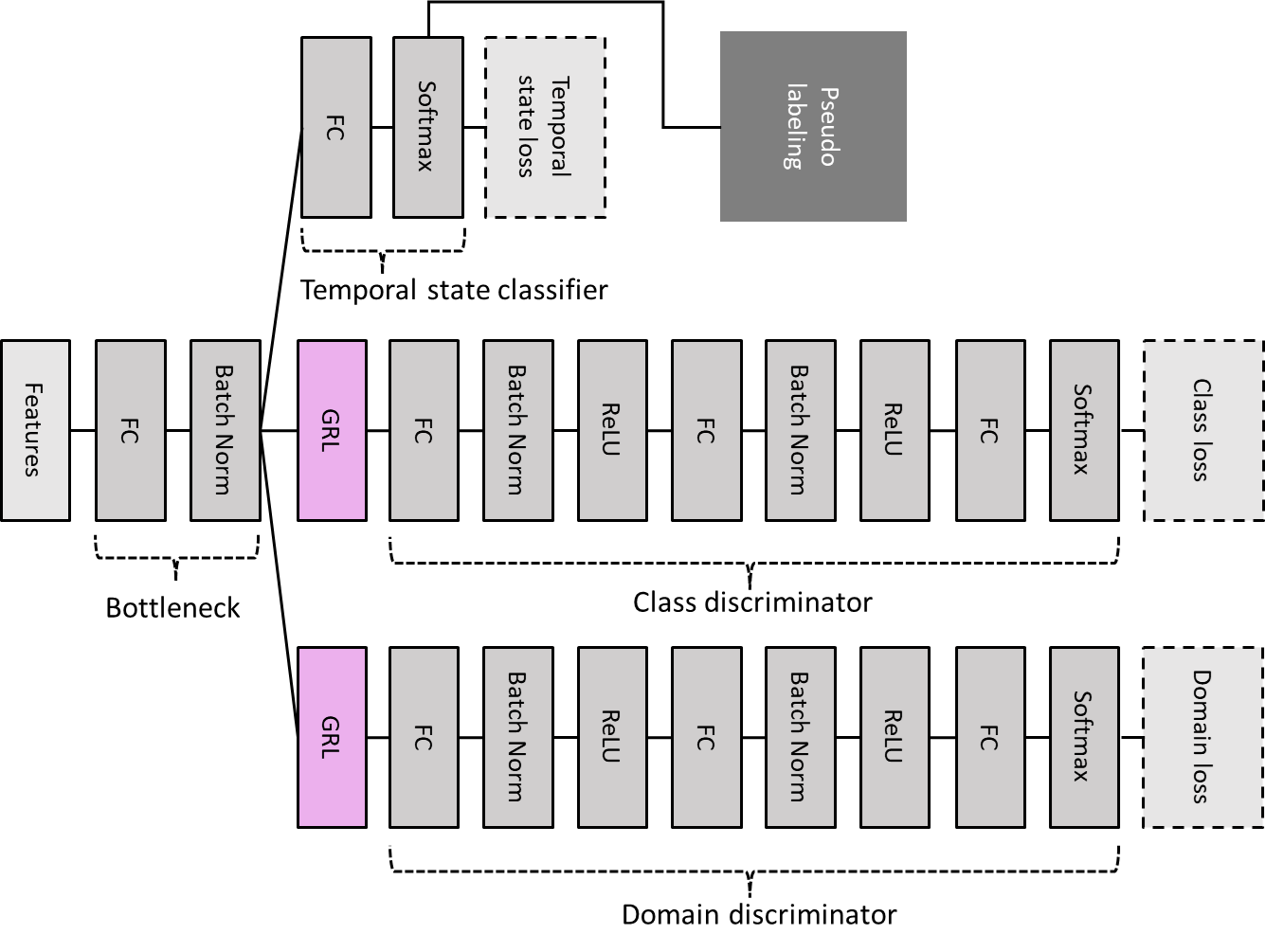}
\caption{The network architecture of user invariant temporal states extraction.\label{User_invariant_temporal_states_extraction}}
\end{figure}

The network architecture of the user-invariant temporal states extraction component is shown in Figure~\ref{User_invariant_temporal_states_extraction}. The extracted features from the feature extractor learned in the previous step are passed through a bottleneck layer, which maps the features through a linear transformation, followed by batch normalization. Subsequent to this feature mapping process, the adjusted features are directed to a temporal state classifier, which employs a linear transformation to categorize them into temporal state classes. (The dark grey rectangle is the Pseudo Temporal State Labeling component that is discussed in detail in the next section.) In parallel, two discriminators are at play: one is tasked with classifying the activity classes and the other with identifying their originating domain. GRL is applied to these two discriminators for adversarial learning. For the discriminator, there are multiple layers. It has two fully connected layers followed by batch normalization and ReLU activations, and finally, a fully connected layer with softmax classification.

Here, we explain the design of loss functions. $\displaystyle h_{bt}$, $\displaystyle h_{ct}^{C} $, $\displaystyle h_{ct}^{D} $, $\displaystyle h_{ct}^{E} $ are the bottleneck, temporal state classifier, class discriminator and domain discriminator of domain-invariant temporal states extraction component respectively. Here, the bottleneck is a feature-mapping function for data distribution adjustment.

Temporal State Loss:
\begin{equation}
L_{tt} =\ l_{CE}\left( h_{ct}^{C}( h_{bt}( h_{f}( x))),\hat{ts}\right)
\end{equation}

Class Loss:
\begin{equation}
 L_{ct} =\ l_{CE}\left( h_{ct}^{D}( R ( h_{bt}( h_{f}( x)))), c \right)
\end{equation}

Domain Loss:
\begin{equation}
L_{dt} =\ l_{CE}\left( h_{ct}^{E}( R (h_{bt}( h_{f}( x)))), d \right)
\end{equation}

The Total Loss:
\begin{equation}
L_{t} = L_{tt} + L_{ct} + L_{dt}
\end{equation}

Where $\displaystyle l_{CE}$ is the cross-entropy loss function. $\displaystyle R$ is the gradient reversal layer.

\subsubsection{Pseudo temporal state labeling}

For the temporal states $\displaystyle \hat{ts} $ annotation, a novel pseudo labeling method which considers user-invariant temporal relation called Pseudo Temporal State Labeling is proposed inspired by \cite{hallac2017toeplitz} as shown in the dark grey rectangle of Figure~\ref{framework_DGTSDA}. For considering user-invariant temporal relation, the temporal consistency property of times series \cite{box2015time} is introduced. The temporal consistency property suggests that data in a time series have an inherent sequential relation, where the observations in a time series don't occur in isolation but are parts of a continuum, where each observation is a progression from the one before it. The temporal consistency property means data that are closer in time are more likely to be similar to each other and belong to the same temporal state than those further apart.

Now, the specific implementation is explained. First, we capture the sub-activities in a deep clustering way\cite{caron2018deep} to learn the initial centroid for each temporal state as follows: 

\begin{equation}
\tilde{u}_{t} =\frac{\sum _{i=0}^{i=N} p_{t}\left( h_{ct}^{C}( h_{bt}( h_{f}( x_{i})))\right) h_{bt}( h_{f}( x_{i}))}{\sum _{i=0}^{i=N} p_{t}\left( h_{ct}^{C}( h_{bt}( h_{f}( x_{i})))\right)}
\end{equation}

Where $\displaystyle \tilde{u}_{t} $ is the $\displaystyle t^{th}$ initial temporal state centroid, $\displaystyle N $ is the number of samples, $\displaystyle p_{t} $ is the value of $\displaystyle t^{th}$ element of the output after the softmax is applied as the dark grey rectangle of Figure~\ref{User_invariant_temporal_states_extraction}. Then,  We use a dynamic programming method with a temporal state switch penalty term to keep the user-invariant temporal consistency property.

\begin{table}[!h]
\caption{The Distance Between Sample and Temporal State Centroid Table.\label{tab_TCD}}
\centering
\resizebox{\columnwidth}{!}{%
\begin{tabular}{l|c|c|c|c|c}
\diagbox{temporal \\ state}{temporal \\ order} & 0 & 1 & 2 & ... & $\displaystyle N-1 $\\
\hline
0 & $\displaystyle D(0,0) $ & $\displaystyle D(1,0) $ & $\displaystyle D(2,0) $ & ... & $\displaystyle D(N-1,0) $\\
\hline
1 & $\displaystyle D(0,1) $ & $\displaystyle D(1,1) $ & $\displaystyle D(2,1) $ & ... & $\displaystyle D(N-1,1) $\\
\hline
... & ... & ... & ... & ... & ... \\
\hline
$\displaystyle T-1 $ & $\displaystyle D(0,T-1) $ & $\displaystyle D(1,T-1) $ & $\displaystyle D(2,T-1)$ & ... & $\displaystyle D(N-1,T-1) $
\end{tabular}%
}
\end{table}

The distance between sample and temporal state centroid is calculated: 

\begin{equation}
D( i,t) =F( h_{bt}( h_{f}( x_{i})),\tilde{u}_{t})
\end{equation}

Here, $\displaystyle i$ is the temporal order index of samples, $\displaystyle F$ is a distance measure function and here we use cosine distance metric. As Table~\ref{tab_TCD} shows, the distance between all the samples and all the temporal state centroids can be calculated, and the distance matrix $\displaystyle M_{D} $ is formed. Then, to encourage temporal neighbouring samples to the same temporal state, a temporal consistency term is applied. When the current temporal state wants to jump to another temporal state in the next time point, the temporal state switch penalty parameter of $\displaystyle \gamma $ is applied to the distance to increase the distance value, while there is no penalty when the current temporal state still stays in the same temporal state at the next time point. As Table~\ref{tab_TCSSPT} shows, a switch penalty matrix $\displaystyle M_{P} $ is set.

\begin{table}[H]
\caption{Temporal State Switch Penalty Table.\label{tab_TCSSPT}}
\centering
\resizebox{\columnwidth}{!}{%
\begin{tabular}{c|c|c|c|c}
\diagbox{current temporal state}{next temporal state} & 0 & 1 & ... & $\displaystyle T-1 $\\
\hline
0 & $\displaystyle 0 $ & $\displaystyle \gamma $ & ... & $\displaystyle \gamma $\\
\hline
1 & $\displaystyle \gamma $ & $\displaystyle 0 $ & ... & $\displaystyle \gamma $\\
\hline
... & ... & ... & ... & ...  \\
\hline
$\displaystyle T-1 $ & $\displaystyle \gamma $ & $\displaystyle \gamma $ & ... & $\displaystyle 0 $
\end{tabular}%
}
\end{table}

Now, the problem is given $\displaystyle M_{D} $ and $\displaystyle M_{P} $, how to find a minimum cost path from time point 0 to $\displaystyle N-1 $. It becomes a classical dynamic programming problem that can be solved.

\newcommand{\algsize}{\fontsize{9.5pt}{9pt}\selectfont}

\begin{algorithm}
\caption{Pseudo temporal state labeling}
\begin{algorithmic}
\algsize 
\REQUIRE \( M_{D} \), an \( T \times N \) matrix representing distances between samples and state centroids for each time step.
\REQUIRE \( M_{P} \), a \( T \times T \) matrix representing penalties for switching between states.
\ENSURE \( path \), an array of size \( N \) representing the sequence of states with the minimum total cost.
\STATE
\STATE Initialize \( future\_costs\_table \) as an \( T \times N \) matrix of zeros
\STATE Initialize \( path \) as an \( N \)-dimensional vector of zeros

\STATE
\STATE \textbf{Backward pass to compute future costs}
\FOR{\( i \leftarrow N-2 \) \TO \( 0 \)}
    \STATE \( j \leftarrow i + 1 \)
    \STATE \( future\_costs \leftarrow future\_costs\_table[:,j] \)
    \STATE \( distances \leftarrow M_{D}[:,j] \)
    \FOR{\( \text{state} \leftarrow 0 \) \TO \( T-1 \)}
        \STATE \( total\_costs \leftarrow future\_costs + distances + M_{P}[:, \text{state}] \)

        \STATE \( future\_costs\_table[i, \text{state}] \leftarrow \min(total\_costs) \)
    \ENDFOR
\ENDFOR
\STATE
\STATE \textbf{Forward pass to determine the best path}

\STATE \( curr\_state \leftarrow \text{argmin}(future\_costs\_table[ :,0] + M_{D}[:,0]) \)
\STATE \( path[0] \leftarrow curr\_state \)

\FOR{\( i \leftarrow 0 \) \TO \( N-2 \)}
    \STATE \( j \leftarrow i + 1 \)
    \STATE \( future\_costs \leftarrow future\_costs\_table[:,j] \)
    \STATE \( distances \leftarrow M_{D}[:,j] \)
    \STATE \( total\_costs \leftarrow future\_costs + distances + M_{P}[:,path[i]] \)
    
    \STATE \( path[i+1] \leftarrow \text{argmin}(total\_costs) \)
\ENDFOR

\RETURN \( \text{path} \)
\end{algorithmic}
\end{algorithm}

The essence of the algorithm is to find the most cost-effective path through a sequence of temporal states over time, considering both the distance costs and the switching penalties. $\displaystyle future\_costs\_table $ is the dynamic programming table that is used to store the computed future costs. Each entry represents the minimum future cost of ending up in a particular state at a particular time step, taking into account the costs of all subsequent decisions. 

The backward pass step involves looking ahead from each state at each time step to determine the minimum cost of reaching the end from that state. This is done in reverse order, from the last time step back to the first. It uses the principle that the best future for any state at any time step can be determined by considering all possible next states and choosing the one with the minimum cost. Once the future costs are known, the algorithm then determines the minimum cost path starting from the first time step. The forward pass step selects the temporal state that minimizes the sum of the distance cost and the future cost at the first time step. For each subsequent time step, the algorithm selects the state that minimizes the sum of the distance cost, the switching penalty from the previous state, and the future cost. The output of the algorithm is the sequence of states that results in the minimum total cost, given the distance costs and the switching penalties.

Next, new centroid of each temporal state is computed based on the updated pseudo temporal state labels: 

\begin{equation}
{u_{t}} = \frac{\sum_{x_i} I(i = t) h_{bt}( h_{f}( x_{i}))}{\sum_{x_i} I(i = t)}
\end{equation}

Where $\displaystyle I(a) $ is an indicator function, returning 1 if $\displaystyle a $ is true and 0 otherwise. Then, the updated pseudo temporal state label is obtained as: 

\begin{equation}
\hat{ts} = \underset{t}{\text{arg min}} F\left( h_{bt}( h_{f}( x_{i})), u_{t}\right)
\end{equation}

\subsection{Cross-user activity classification}
This component aims to learn a user-invariant activity classifier to achieve the the goal of enhanced cross-user HAR. After the user-invariant temporal states are derived from the component (2), these temporal states can act as an intermediate representation that bridges the variability between users. These labels provide prior knowledge for the cross-user activity classifier. Since they are user-invariant, they provide an abstract view of activity sequences that hold true across diverse users. The temporal state labels act like a regularization to reduce the risk of overfitting to specific users for constraining the cross-user HAR process. They offer a consistent set of patterns that the activity classifier can reliably refer to, making the task of recognizing activities across different users more effective and accurate.

Moreover, adversarial learning also happens between component (1) and this component. Component (1) acts like discriminator in GAN domain adaptation that discerns the differences by identifying how different users are represented in the sensor data. This component acts like generator in GAN domain adaptation that attempts to confuse the data distributions between source and target users. In the same time, this component also aims to identify activity labels from source user. The back-and-forth between components (1) and this component drives both processes to continually refine and improve, leading to a user-invariant activity classification for a more robust and effective cross-user HAR model.

Within this component, the adversarial learning GRL technique is applied to adjust the feature distribution that is not only preserve the user-invariant temporal relation knowledge but also ensuring the activity decision boundaries are general across users. As shown in Figure~\ref{framework_DGTSDA} (3), the temporal state loss aims to correctly categorize the temporal states, and the source user class loss is designed for activity classification. Simultaneously, the domain loss strives to confuse the domain discriminators. This
confusion forces the learning process to focus on learning user-invariant decision boundaries of activity classes as shown in the example in Figure~\ref{DTSDA_schematic_diagram} (3).

\begin{figure}[h!]
\centering
\includegraphics[width=\columnwidth]{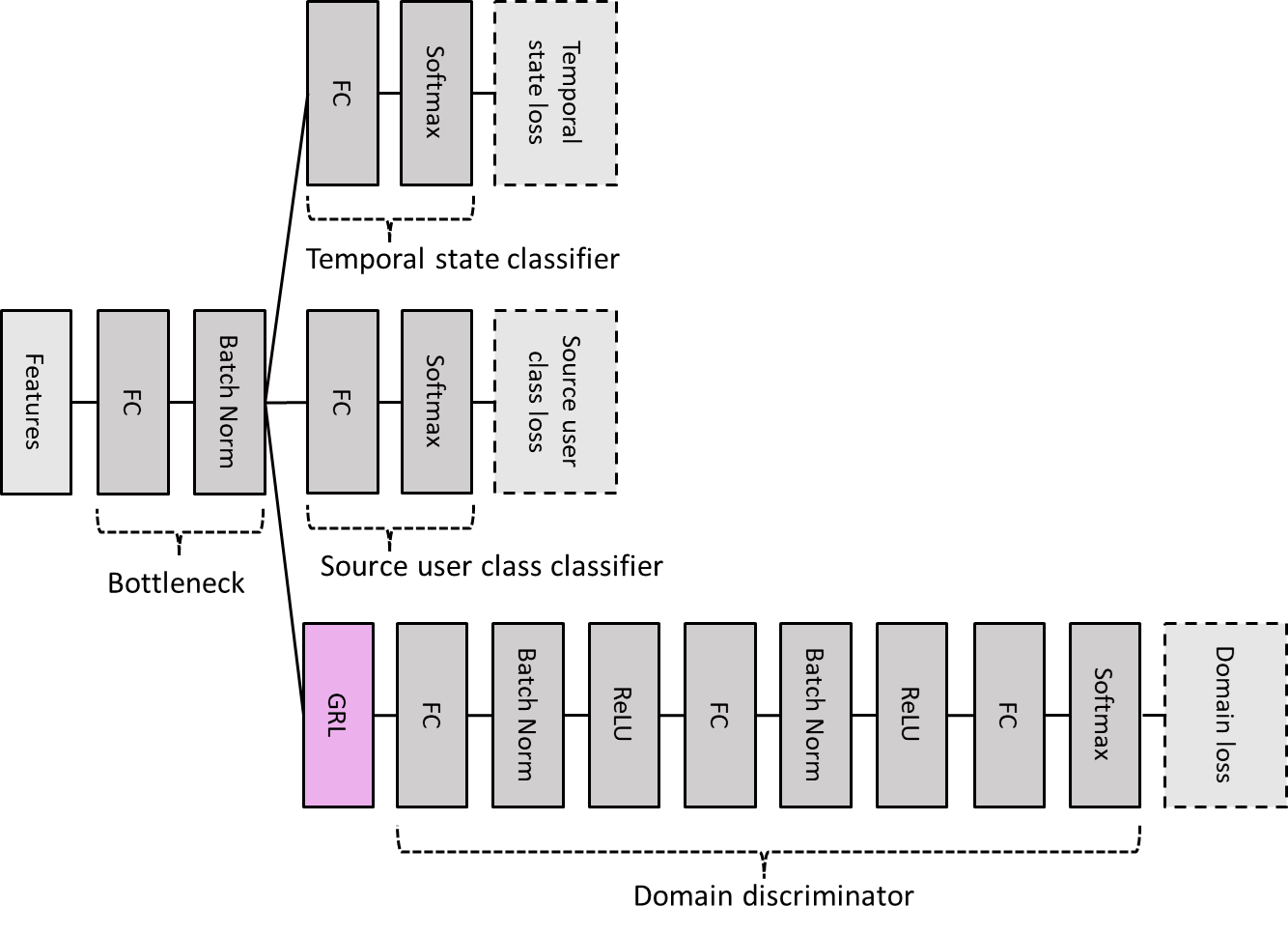}
\caption{The network architecture of cross-user activity classification.\label{User_invariant_classification}}
\end{figure}

The network architecture of the cross-user activity classification component is shown in Figure~\ref{User_invariant_classification}. The network takes extracted features from feature extractor and adjust the distribution using the bottleneck, and then followed by three parallel processes. The first process aims to classify temporal state for temporal relation learning; the second process aims to classify source user activity classes. the last process aims to learn user-invariant representation.

Here, we explain the design of loss functions. $\displaystyle h_{bc} $, $\displaystyle h_{cc}^{F} $, $\displaystyle h_{cc}^{G} $, $\displaystyle h_{cc}^{H} $ are the bottleneck, temporal state classifier, source user class classifier and domain discriminator of cross-user activity classification component, respectively.

Temporal State Loss:
\begin{equation}
L_{tc} = l_{CE}\left( h_{cc}^{F}( h_{bc}( h_{f}( x))),\hat{ts}\right)
\end{equation}

Source User Class Loss:
\begin{equation}
L_{cc} =\ l_{CE}\left( h_{cc}^{G}(h_{bc}( h_{f}( x_{source}))), c_{source} \right)
\end{equation} 

Domain Loss:
\begin{equation}
L_{dc} =\ l_{CE}\left( h_{cc}^{H}( R (h_{bc}( h_{f}( x)))), d \right)
\end{equation}

The Total Loss:
\begin{equation}
L_{c} = L_{tc} + L_{cc} + L_{dc}
\end{equation}

Where $\displaystyle l_{CE}$ is the cross-entropy loss function. $\displaystyle R $ is the gradient reversal layer. Here, only the labelled source user's data is used because of the unknown labels of the target user.

\begin{figure}[H]
\centering
\includegraphics[width=0.8\columnwidth]{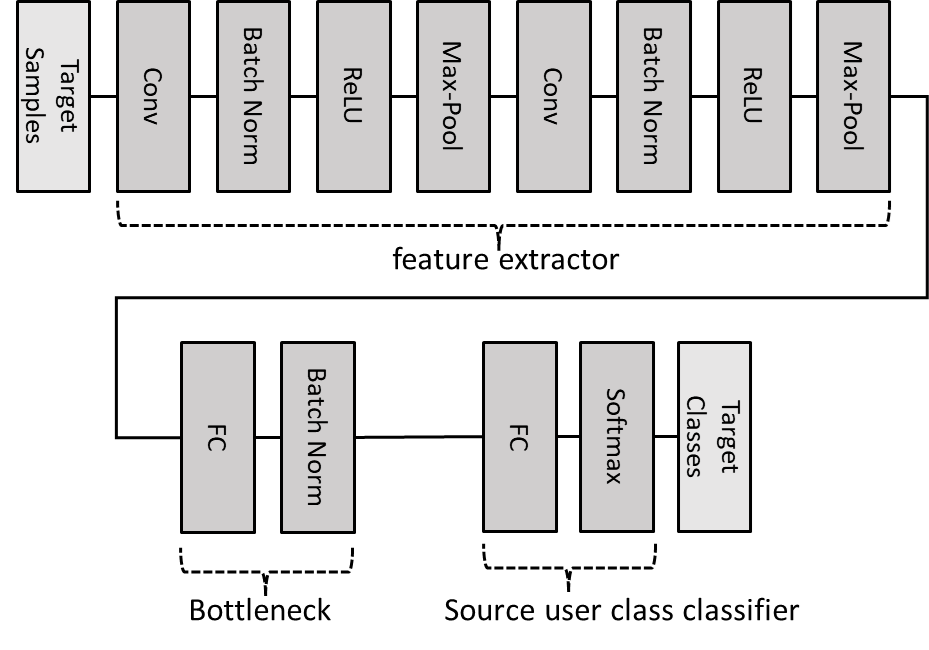}
\caption{The network architecture of target user classification.\label{target_process}}
\end{figure}

The overall learning process of the DTSDA framework is inspired by the work in \cite{lu2022out}, that iteratively trains the adversarial components. After the training process, the bottleneck and source user classifier in the cross-user activity classification component combined with the global feature extractor module is the final model that we need for target user classification as shown in Figure~\ref{target_process}.

\section{Experiments}

In this section, we evaluate the performance of DTSDA via extensive experiments on cross-user activity recognition.

\subsection{Datasets and pre-processing}
We used three common sensor-based human activity recognition public datasets for validating our cross-user HAR method. Table~\ref{tab_datasets_info} shows the subjects and the common activities among the users in each dataset. Here, we only use the sensor values of the accelerometer and gyroscope in the position of the right lower arm to explore the practical scenario of using the smartwatch. In the following, the information of each dataset is introduced briefly.

\begin{table}[!h]
\caption{Three sensor-based HAR datasets information Table.\label{tab_datasets_info}}
\centering
\resizebox{\columnwidth}{!}{%
\begin{tabular}{|p{1cm}|p{1cm}|p{1.2cm}|p{5.8cm}|}
\textbf{Dataset \& Ref.} & \textbf{Subjects} & \textbf{\#Activities} & \textbf{Common Activities} \\ \hline
OPPT & S1, S2, S3, S4 & 4 & lying, sitting,   standing, walking \\ \hline
PAMAP2 & 1, 2, 3, 4, 5, 6, 7, 8, 9 & 11 & \begin{tabular}[c]{@{}l@{}}lying, sitting, standing, walking, \\ running, cycling, Nordic walking, \\ ascending stairs, descending stairs, \\ vacuum cleaning, ironing\end{tabular} \\ \hline
DSADS & 1, 2, 3, 4, 5, 6, 7, 8 & 19 & \begin{tabular}[c]{@{}l@{}}sitting, standing, lying on back, \\ lying on right, ascending stairs, \\ descending stairs, standing in an elevator still, \\ moving around in an elevator, \\ walking in a parking lot, \\ walking on a treadmill in flat, \\ walking on a treadmill inclined positions, \\ running on a treadmill in flat, \\ exercising on a stepper, \\ exercising on a cross trainer, \\ cycling on an exercise bike in horizontal positions, \\ cycling on an exercise bike in vertical positions, \\ rowing, jumping, playing basketball\end{tabular}
\end{tabular}%
}
\end{table}

\textbf{OPPORTUNITY Dataset (OPPT) \cite{chavarriaga2013opportunity}:}
The OPPT dataset captures subjects in a daily living scenario performing morning activities. Unlike more structured data collection approaches, subjects in this dataset perform activities based on a loose description, introducing a natural variability in how each user is executed. This mirrors realistic scenarios where activities are not strictly choreographed, making it valuable for assessing the real-world applicability of HAR models.

\textbf{Physical Activity Monitoring Dataset (PAMAP2)\cite{reiss2012introducing}:}
PAMAP2, while having a protocol, does not impose a highly restrictive set of guidelines on its subjects. Instead, the protocol serves as a general introduction, allowing for activity variability for each user. With over 10 hours of cumulative data, this dataset represents scenarios where users might have a basic understanding of the activities but still perform them with personal variations.

\textbf{Daily and Sports Activities Dataset (DSADS) \cite{barshan2014recognizing}:}
DSADS stands out for its emphasis on capturing activities as they naturally occur. Subjects perform activities without specific instructions, leading to inter-subject variations. With 19 activity classes, including closely related ones like three different walking activities, DSADS offers the most challenging domain adaptation scenario among the three datasets.

The selection of these three datasets provides a gradient of cross-user HAR challenges. Starting with OPPT's 4 activities, to PAMAP2's 11, and culminating with DSADS's 19, they collectively offer increasing complexity. By evaluating models across these datasets, we ensure a comprehensive assessment of their adaptability, generalization and robustness. Furthermore, the datasets differ in their collection methodologies. Evaluating these datasets allows us to understand model performance in more practical scenarios.

For data segmentation, a sliding window technique, commonly used in sensor-based HAR, is employed. Given the focus on temporal relation extraction, each window is set to 3 seconds with a 50\% overlap as a common setting in sensor-based HAR \cite{wang2018impact}.

\subsection{Experimental setup}

\begin{figure}[H]
\centering
\includegraphics[width=0.55\columnwidth]{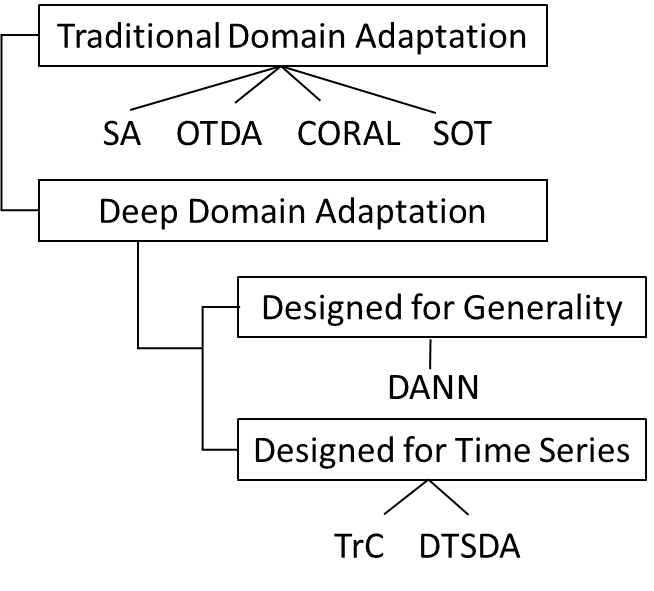}
\caption{The categories of methods for comparison.\label{categories_of_methods}}
\end{figure}

We adopt six comparison methods categorized as shown in Figure~\ref{categories_of_methods}, traditional domain adaptation and deep domain adaptation methods that target specifically time series data or data agnostic. Moreover, TrC needs limited target domain labels for fine-tuning, while other methods do not.

\textbf{SA} \cite{fernando2013unsupervised}: is a technique that is proposed to align the subspaces of source and target data so that they become more similar.

\textbf{OTDA} \cite{flamary2016optimal}: aims to transport one probability distribution to another in an optimal way for mapping source domain samples to target domain samples inspired by optimal transport theory.

\textbf{CORAL} \cite{sun2016return}: is a technique that leverages the alignment of covariance within feature layers to foster enhanced domain-invariant characteristics.

\textbf{SOT} \cite{lu2021cross}: explores the substructure of domains to complete substructure-level mapping to achieve a balance between coarse-grained mapping and fine-grained mapping.

\textbf{DANN} \cite{ganin2016domain}: is a deep domain adaptation technique that employs adversarial training to compel the discriminator to fail at distinguishing between domains, thereby fostering superior domain-invariant features.

\textbf{TrC} \cite{rokni2018personalized}: is a deep domain adaptation method that uses CNN to capture temporal features. The fine-tuning framework is applied to the model trained from source user data.

For the above methods, hyper-parameters tuning is implemented, and we follow a
similar methodology as in \cite{flamary2016optimal} to prevent overfitting on the testing set. The target user data is divided into two distinct sets: validation and test sets. The validation set's role is to achieve the best possible accuracy by fine-tuning the range of hyper-parameters. Upon determining the optimal hyper-parameters, we assess the model's performance using the test set. The chosen metric for evaluation is the classification accuracy of the target user.

\subsection{Classification results}
We first analyse the classification results of different methods. These experiments are conducted on OPPT, PAMAP2, and DSADS datasets. The performance of each method is evaluated based on the transition from one user to another user. Three users are randomly selected from all the users in each dataset. Then, a one-to-one cross-user HAR task is implemented, encompassing all possible pairings between users in each dataset.

\begin{figure}[h!]
\centering
\includegraphics[width=\columnwidth]{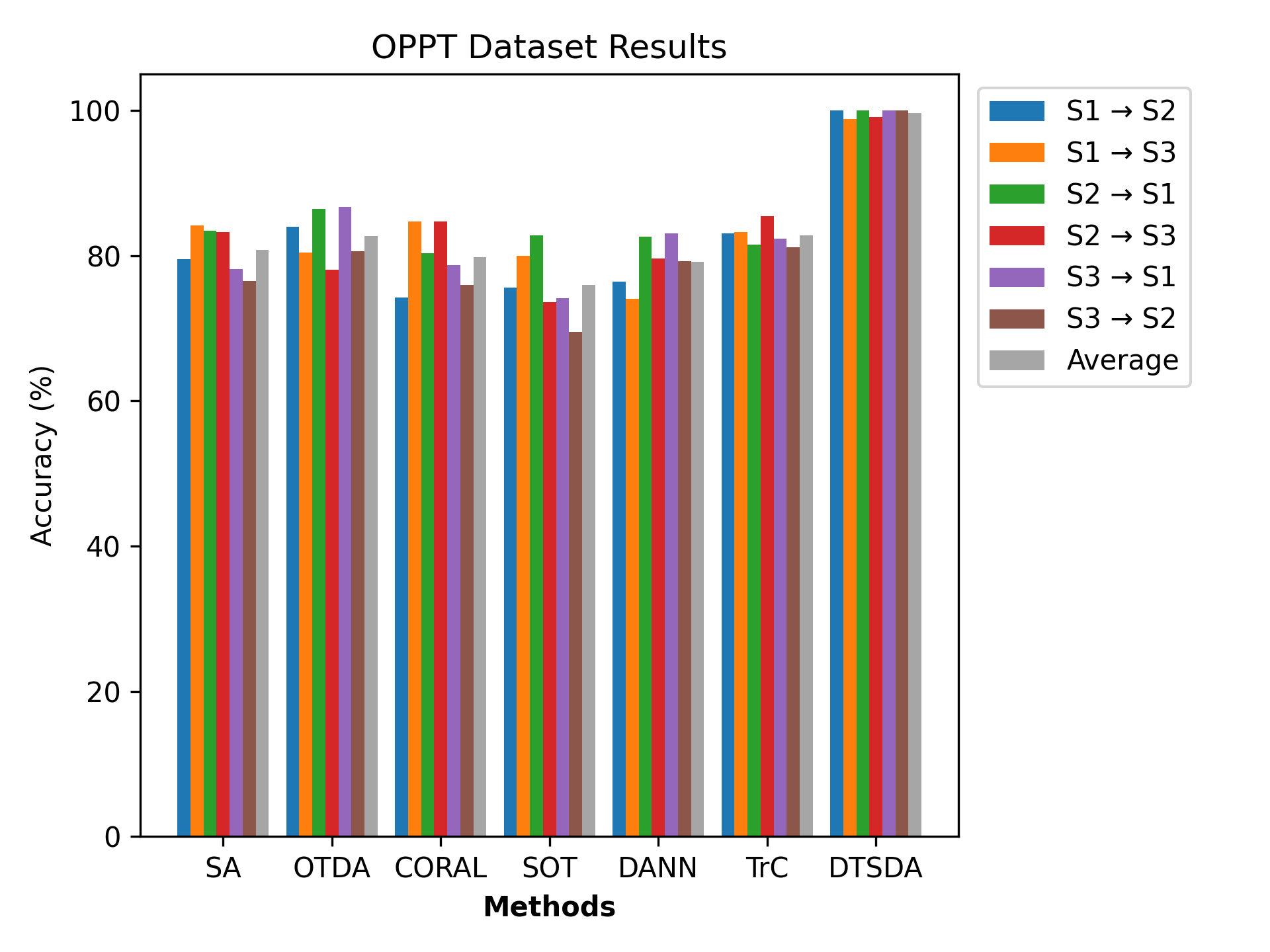}
\caption{OPPT dataset classification results.\label{OPPT_Dataset_Results}}
\end{figure}

On the OPPT dataset (see Figure~\ref{OPPT_Dataset_Results}), DTSDA outperformed all other methods by a significant margin, achieving perfect or near-perfect results in all scenarios. SA, OTDA and TrC are competitive in their performance, with TrC generally outperforming SA in most transfers, especially in S3 to S2 and S3 to S1 transfers. DANN, CORAL and SOT show more variable results, performing relatively weaker in some tasks, indicating potential limitations in certain domain adaptation scenarios.

\begin{figure}[h!]
\centering
\includegraphics[width=\columnwidth]{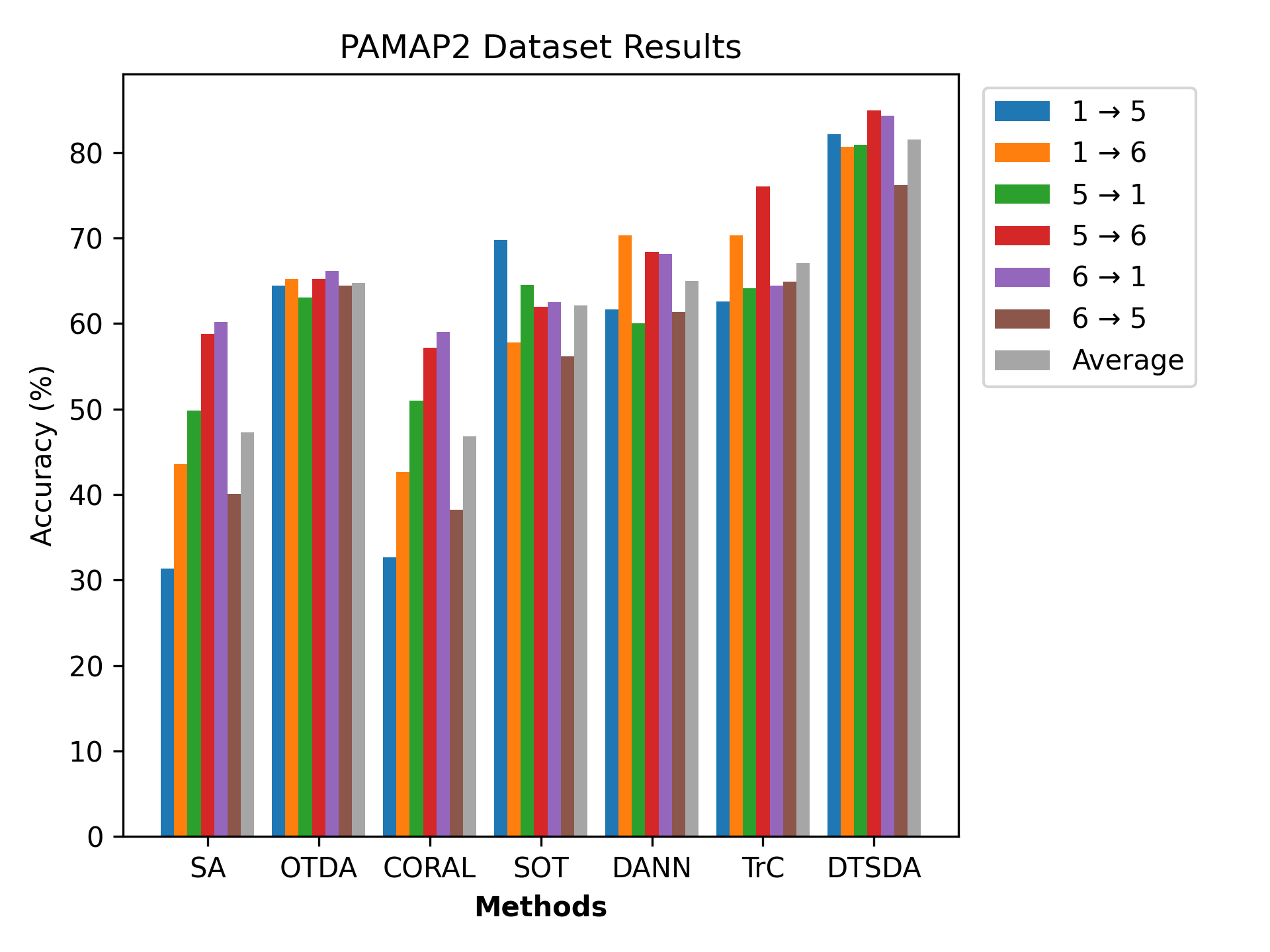}
\caption{PAMAP2 dataset classification results.\label{PAMAP2_Dataset_Results}}
\end{figure}

On the PAMAP2 dataset (see Figure~\ref{PAMAP2_Dataset_Results}), DTSDA again performed the best in all scenarios, with results ranging from 78.21\% to 87.95\%. OTDA, SOT, DANN and TrC appear to be the next tier in terms of performance, with OTDA being more consistent across various tasks while TrC excels particularly in the 1 to 6 and 5 to 6 transfers. DANN performs well in some scenarios like 1 to 6 and struggles in others like 1 to 5. SA and CORAL remain consistently lower than the other methods, implying potential limitations in domain adaptation capabilities for the PAMAP2 Dataset.

\begin{figure}[h!]
\centering
\includegraphics[width=\columnwidth]{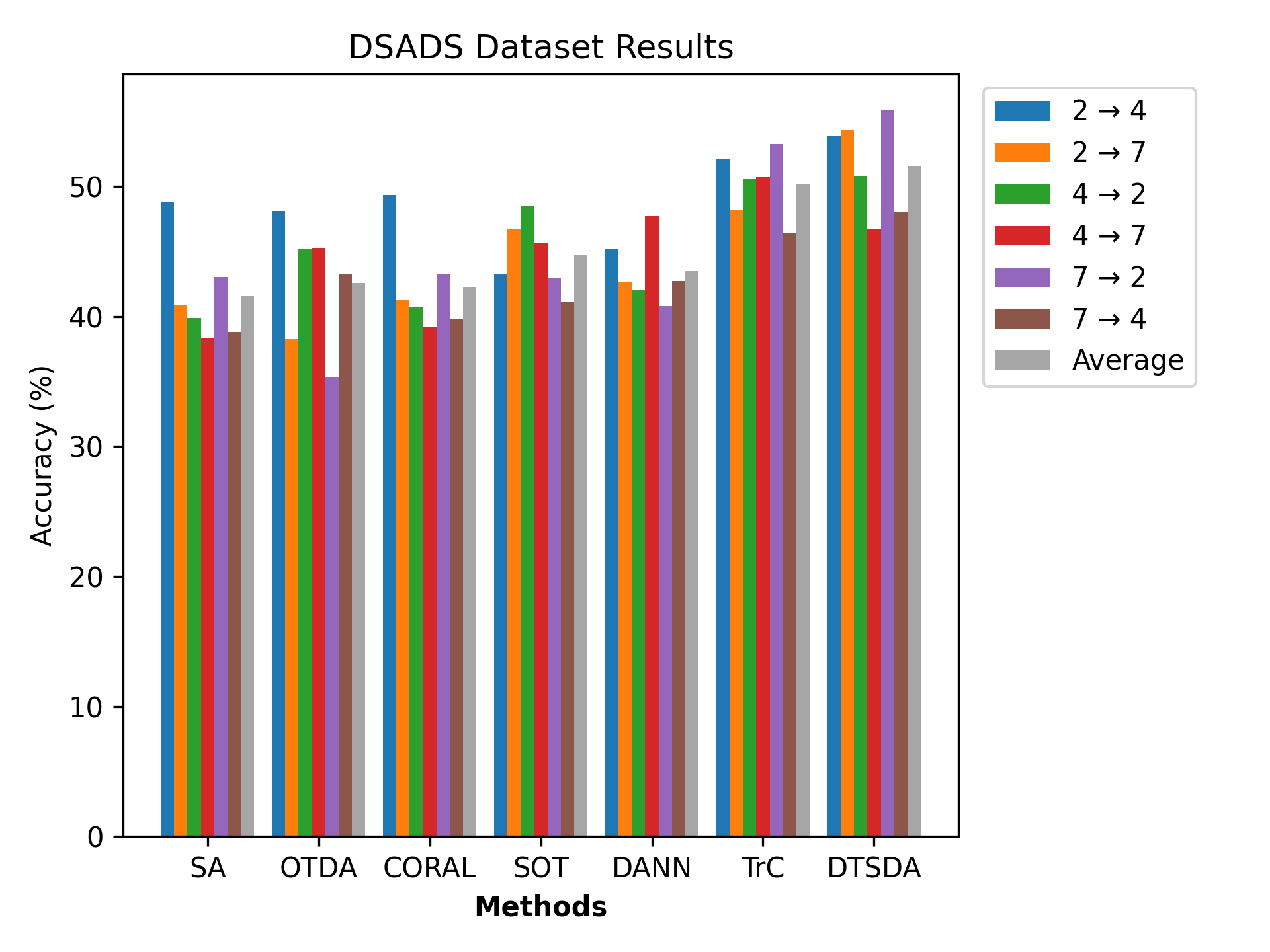}
\caption{DSADS dataset classification results.\label{DSADS_Dataset_Results}}
\end{figure}

On the DSADS dataset (see Figure~\ref{DSADS_Dataset_Results}), the results were more varied. DTSDA still performed the best overall, but not by as large a margin as with the other datasets. In particular, TrC achieved similar results in several scenarios. The rest of the methods, including SA, OTDA, CORAL, SOT, and DANN, show mixed results with no clear patterns of dominance across the transfers.

\begin{table}[htbp]
\centering
\caption{Average accuracies and standard deviations for each method across different datasets.}
\setlength{\tabcolsep}{2pt} 
\renewcommand{\arraystretch}{1.2} 
\begin{tabular}{|l|l|l|l|l|l|l|l|l|}
\hline
Dataset & Metrics & SA & OTDA & CORAL & SOT & DANN & TrC & DTSDA \\ \hline
\multirow{2}{*}{OPPT} & AA & 80.82 & 82.69 & 79.77 & 75.92 & 79.16 & 82.78 & 99.65 \\ \cline{2-9}
                      & SD & 2.9 & 3.22 & 4.02 & 4.33 & 3.17 & 1.41 & 0.5 \\ \hline
\multirow{2}{*}{PAMAP2} & AA & 47.3 & 64.76 & 46.78 & 62.12 & 64.98 & 67.08 & 81.55 \\ \cline{2-9}
                         & SD & 10.21 & 0.94 & 9.69 & 4.47 & 4.06 & 4.68 & 2.87 \\ \hline
\multirow{2}{*}{DSADS} & AA & 41.63 & 42.57 & 42.27 & 44.7 & 43.51 & 50.21 & 51.59 \\ \cline{2-9}
                        & SD & 3.57 & 4.42 & 3.42 & 2.49 & 2.31 & 2.28 & 3.36 \\ \hline
\end{tabular}
\label{tab_accuracy_std}

* AA: Average Accuracy (\%), SD: Standard Deviation of Accuracy (\%)
\end{table}

Table~\ref{tab_accuracy_std} is the average accuracies and standard deviations for each method across OPPT, PAMAP2 and DSADS datasets. For the OPPT dataset, DTSDA significantly outperformed other methods with an impressive average accuracy of 99.65\% and the lowest standard deviation of 0.5\%. This suggests a high level of consistency and reliability in its performance. The TrC method also showed strong results with an AA of 82.78\% and a low SD of 1.41\%, indicating its effectiveness in this context. Traditional methods like SA, OTDA, and CORAL demonstrated moderate accuracies ranging from 79.77\% to 82.69\%, but with higher variability as indicated by their SD values.

In the PAMAP2 dataset, DTSDA again led with a high accuracy of 81.55\%, suggesting its robustness across different datasets. The next best performing method was TrC with an AA of 67.08\%. The substantial gap in performance between DTSDA and other methods, including TrC, indicates a notable superiority of DTSDA in handling the challenges presented by the PAMAP2 dataset. The methods OTDA, SOT and DANN showed moderate performances, with accuracies of more than 60\%, but OTDA has the lowest standard deviations implying more consistency.

The DSADS dataset presented a more challenging scenario for all methods, as evidenced by overall lower accuracy scores. DTSDA and TrC have the AAs of 51.59\% and 50.21\%, respectively, but these values are significantly lower than those observed in the other datasets. This suggests that DSADS dataset may have inherent complexities or characteristics that make accurate predictions more challenging for these methods.

In summary, DANN method, despite being a deep learning method, did not consistently outperform the traditional domain adaptation methods, including SA, OTDA, CORAL and SOT, which indicates there are no guarantees that deep domain adaptation is more suitable for time series domain adaptation tasks. TrC is designed for time series data, uses deep learning, and uses target domain labels. These characteristics make it well-suited for tasks involving time series data and might explain its relatively high performance, especially in complex tasks like in the DSADS dataset. DTSDA, like TrC, is designed for time series data and uses deep learning. However, it doesn't use target domain labels. Despite this, it consistently outperforms other methods. This could be due to a more effective implementation that makes use of temporal relation knowledge. In summary, across all three datasets, DTSDA consistently performed the best, while the performance of the other methods was more varied and dependent on the specific scenario. This suggests that DTSDA is a robust method for cross-user HAR task, performing well across different datasets and scenarios.

\subsection{Effect of temporal relation knowledge}

In this section, we analyse the effect of temporal relation knowledge. The experiments are performed on OPPT, PAMAP2 and DSADS datasets. SOT, DANN and DTSDA methods are compared. SOT is a state-of-the-art traditional domain adaptation method, and DANN is a deep domain adaptation method. Both of the methods are not designed for time series domain adaptation compared to our DSADS method.

\begin{table}[htbp]
\centering
\caption{Activity class and index mapping table of the three datasets.}
\setlength{\tabcolsep}{6pt} 
\renewcommand{\arraystretch}{1} 
\begin{tabular}{lrl}
\toprule
Dataset &  Index &                                         Activity \\
\midrule
   OPPT &      1 &                                         standing \\
        &      2 &                                          walking \\
        &      3 &                                          sitting \\
        &      4 &                                            lying \\
 \midrule
 PAMAP2 &      1 &                                            lying \\
        &      2 &                                          sitting \\
        &      3 &                                         standing \\
        &      4 &                                          walking \\
        &      5 &                                          running \\
        &      6 &                                          cycling \\
        &      7 &                                   Nordic walking \\
        &      8 &                                 ascending stairs \\
        &      9 &                                descending stairs \\
        &     10 &                                  vacuum cleaning \\
        &     11 &                                          ironing \\
 \midrule
  DSADS &      1 &                                          sitting \\
        &      2 &                                         standing \\
        &      3 &                                    lying on back \\
        &      4 &                                   lying on right \\
        &      5 &                                 ascending stairs \\
        &      6 &                                descending stairs \\
        &      7 &                       standing in elevator still \\
        &      8 &                        moving around in elevator \\
        &      9 &                           walking in parking lot \\
        &     10 &                     walking on treadmill in flat \\
        &     11 &          walking on treadmill inclined positions \\
        &     12 &                     running on treadmill in flat \\
        &     13 &                            exercising on stepper \\
        &     14 &                      exercising on cross trainer \\
        &     15 & cycling on exercise bike in horizontal positions \\
        &     16 &   cycling on exercise bike in vertical positions \\
        &     17 &                                           rowing \\
        &     18 &                                          jumping \\
        &     19 &                               playing basketball \\
\bottomrule
\end{tabular}
\label{tab_activity_index_table}
\end{table}

\begin{figure}[h!]
\centering
\includegraphics[width=\columnwidth]{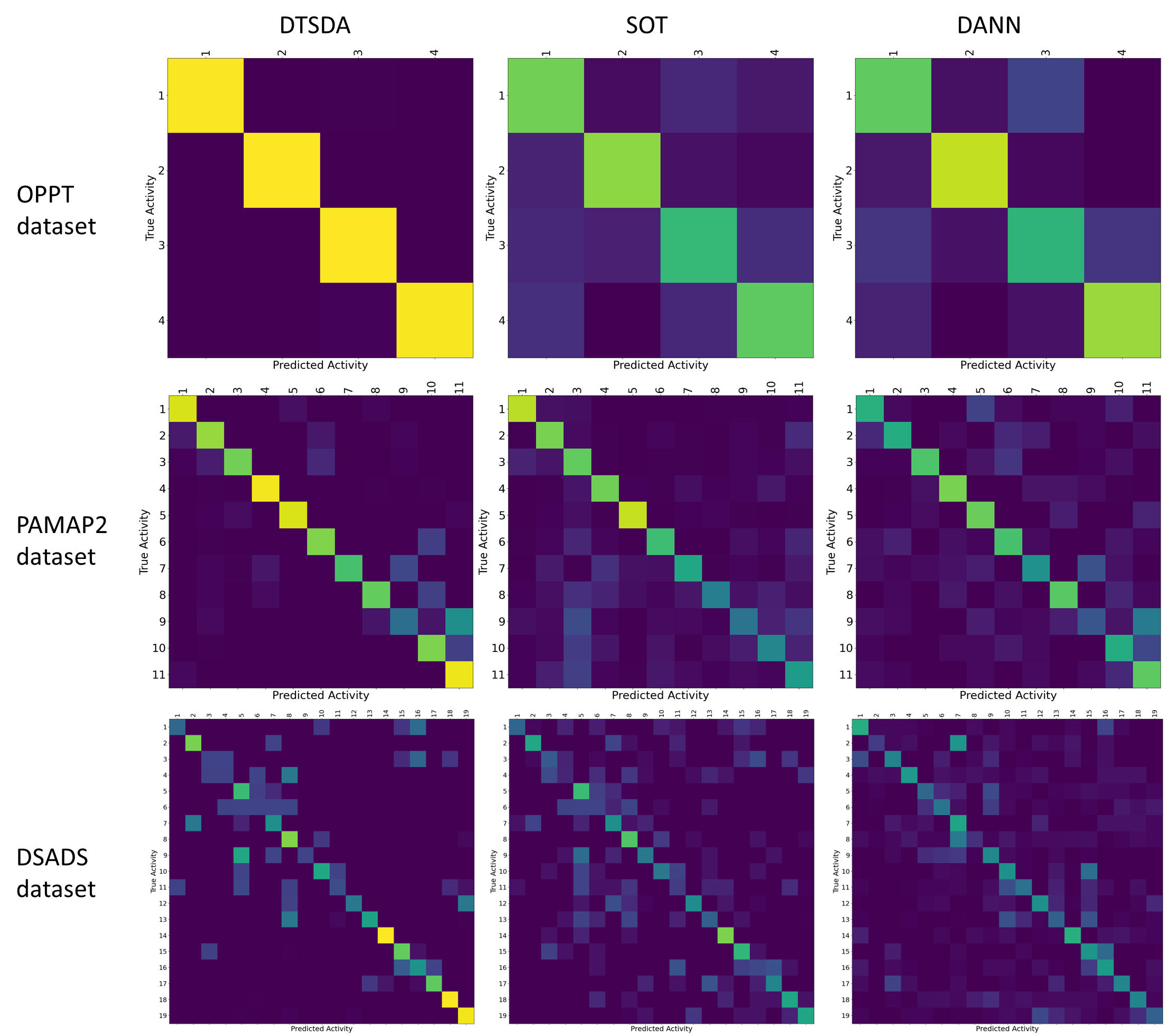}
\caption{Confusion matrices of DTSDA, SOT and DANN methods for the average performance of tasks in the DSADS, PAMAP2 and OPPT datasets.\label{cm_results}}
\end{figure}

Table~\ref{tab_activity_index_table} is the activity class and index mapping table of the three datasets. Figure~\ref{cm_results} is the confusion matrices of DTSDA, SOT and DANN methods for the average performance of tasks in the DSADS, PAMAP2 and OPPT datasets. The warmer the colour (closer to yellow), the higher the value, and vice versa. The x-axis represents the predicted activities, while the y-axis represents the true activities. The diagonal from the top left to bottom right represents correct predictions (true positives), where the predicted activity matches the true activity.

All methods tend to have obvious values along the diagonal. This indicates that the majority of activities are recognized correctly by all the methods, reflecting their effectiveness in cross-user activity recognition. All methods can identify activities such as 'lying', 'sitting', 'standing', 'running' and 'cycling' efficiently. DTSDA can further identify 'ironing', 'vacuum cleaning', 'walking', 'ascending stairs', and 'Nordic walking' with higher accuracy compared to the other methods. Furthermore, the SOT and DANN method indicates a dispersed classification outcome, while DTSDA has less scattered and more clarity in the classification because of the temporal state and sub-activity extraction.

Activities that have distinct characteristics, such as "exercising on cross trainer", are recognized well by all methods. Activities like 'lying', 'walking', and 'running' have relatively high accuracy in all the methods. This could be because these activities have distinct movement patterns that are easily distinguishable from other activities. For example, 'lying' probably has minimal motion; walking and running activities inherently vary by pace, rhythm, and gait. Walking is more rhythmic and slower, while running is faster-paced with longer strides. The consistent and rhythmic pattern of walking and running is characterized by repeated heel-to-toe sequences and a generally steady pace. This suggests that unique activities, regardless of the method used, can be detected with high accuracy. 

All methods show confusion between activities that are inherently similar to each other. For example, three types of walking activity (i.e. 'walking in parking lot', 'walking on treadmill in flat', and 'walking on treadmill inclined positions') tend to be mixed up. "sitting" gets misclassified as "cycling" activities because the arms are static in both cases. Some activities have similar temporal patterns, making them harder to distinguish. For example, 'ascending stairs' and 'descending stairs' both involve foot movements with periodic hand movements (if one is holding a railing). The similarity arises in the gait and hand motion. The reason could be the sensor values are only used from the right lower arm which lacks foot movement information. 'lying on the back' and 'lying on the right' both involve lying down, with only the orientation differing. The temporal patterns for changing postures might be similar, causing misclassification for DTSDA. For the DANN, orientation might be a distinguishing factor that leads to a better performance. 

SOT and DANN methods can be confused with low-motion activities like 'standing', 'sitting' and 'lying', while DTSDA can identify these activities very well. 'Lying' typically involves an extended period of inactivity, which may be captured as a consistent pattern over time windows, while 'sitting' and 'standing' might show more variation in the sensor data due to occasional adjustments in posture. 'Standing' and 'sitting' might have overlapping features that both involve periods of stillness, which could confuse models that heavily rely on motion-based features. DTSDA can capture the temporal relation context and subtle variations in sensor readings over time that differentiate between similar low-motion activities. That may be the reason for the better performance of the DTSDA method compared to the other methods. 

In addition, DTSDA seems to better differentiate dynamic activities, like "standing in elevator still" vs "moving around in elevator", more effectively. Moreover, DTSDA seems to excel in identifying activities with complex temporal relations, such as 'playing basketball', 'jumping' and 'rowing'. Playing basketball is dynamic with varied motions. The diverse motions create a rich temporal pattern. Jumping has a distinct motion with periods of aerial time. The repeated pattern of leaving and hitting the ground. Rowing is a cyclic motion with consistent force exertion. The rhythmic pulling offers a recognizable pattern. This suggests that activities with changing temporal patterns can be recognized better using DTSDA.

Activities might include transitional movements that aren't strictly part of the activity but happen frequently enough to be considered. For instance, while 'ironing', people might periodically adjust the garment. 'vacuum cleaning' may include some vacuum machine position movement sub-activity. These adjustments might resemble other activities, leading to confusion. Compared to SOT and DANN, DTSDA captures the temporal relation knowledge that may lead to a better understanding of the transitional movements.

In summary, the utilization of temporal relation knowledge shows the function of reducing the distribution differences between source and target users, especially for dynamic activities with complex temporal relations. The extraction of temporal relation knowledge embedded in time series data enhances the robustness and accuracy of activity recognition. Leveraging the inherent sequences and patterns in time-ordered data allows for a more nuanced understanding of various activities, especially when activities possess rhythmic or sequential characteristics. Thus, the incorporation of temporal knowledge not only augments the recognition capabilities but also offers deeper insights into the intricacies of different human activities.

\section{Conclusion}

This study presented the DTSDA method, a novel approach to time series domain adaptation. Our innovative approach diverges from traditional models by incorporating temporal relations over time rather than assuming samples independent of each other. By employing the adversarial learning strategy and introducing the concept of temporal state, DTSDA is able to perform cross-user HAR effectively. Furthermore, the Pseudo Temporal State Labeling technique further enhances the performance for extracting user-invariant temporal states.

DTSDA's efficacy was demonstrated through experiments on three publicly available activity recognition datasets, where it outperformed other models in a cross-user HAR scenario. These promising results highlight the potential of our approach, presenting a new method in domain adaptation for HAR and opening new avenues for future research. The effect of temporal relation knowledge is explored. The adoption of temporal relation knowledge in domain adaptation could bridge the gap between source and target users in the HAR field.

\bibliographystyle{IEEEtran}
\bibliography{ref}

\begin{IEEEbiography}[{\includegraphics[width=1in,height=1.25in,clip,keepaspectratio]{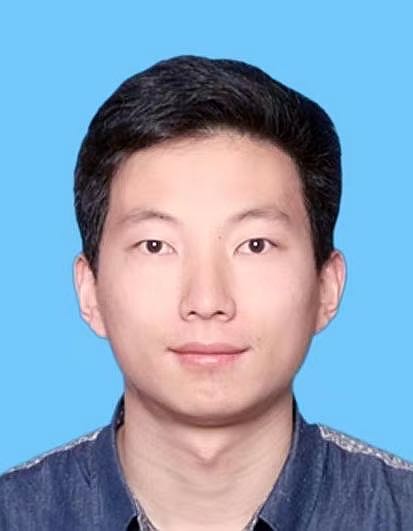}}]{Xiaozhou Ye}
received the M.Sc. degree from Hohai University, Nanjing, China, in 2017. He is currently
pursuing the Ph.D. degree with the Department of Electrical, Computer, and Software Engineering, The University of Auckland, Auckland, New
Zealand.

He worked in industries for business intelligence and data analysis in New Zealand from 2018 to 2020. His current research interests include transfer learning, human activity recognition
(HAR), and pervasive healthcare systems.
\end{IEEEbiography}

\begin{IEEEbiography}[{\includegraphics[width=1in,height=1.25in,clip,keepaspectratio]{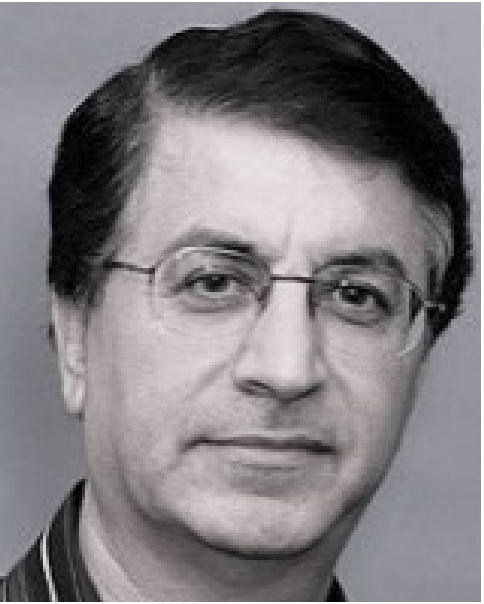}}]{Waleed H. Abdulla}(Senior Member, IEEE) received the Ph.D. degree from the University of Otago, New Zealand, in 2002.

He is currently an Associate Professor with The University of Auckland, New Zealand. He coauthored a book in audio watermarking, which has been downloaded over 8763 times. The book includes many topics in the psychology of hearing. His research activities are under 'signal processing, analysis, and recognition,' which encompass multidisciplinary topics. He focuses on fundamental and applied research in domains with direct communal relevance including, human health and well-being and economic impact. His specific recent research activities with Ph.D. students are in: speech enhancement, speaker recognition, speech processing, hyperspectral imaging for detecting honey botanic origins, diagnosing diabetic retinopathy, active noise control, human biometrics, lase like ultrasound signal communication. He supervised over 30 Ph.D. and master’s degree students.

Mr. Abdulla served as the Vice President Member of Relations and Development in Asia Pacific Signal and Information Processing Association (APSIPA) for two terms followed by two-year Board of Governors. He is one of the steering committee members who established APSIPA, in 2009. His awards and honors include the APSIPA Distinguished Lecturer Award and The University of Auckland Faculty Best Teachers of the Year in 2005 and 2012. He won two best paper awards in 2012 and 2016 in two major conferences. He is the APSIPA Newsletter Founder and was the Editorin-Chief. He is the General Chair of APSIPA 2020 Conference. He visited and delivered talks in several universities and conferences as a Presenter and Keynote Speaker. He has been serving as an editorial board member for five journals.
\end{IEEEbiography}

\begin{IEEEbiography}[{\includegraphics[width=1in,height=1.25in,clip,keepaspectratio]{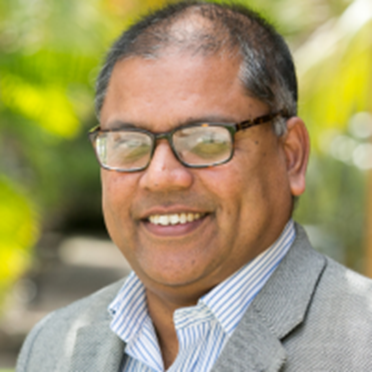}}]{Nirmal Nair} (S’89-M’04-SM’10) received BE in Electrical Engineering from Maharaja Sayajirao University, Baroda, 1990 and ME in High Voltage from Indian Institute of Science, Bangalore, 1996. After a decade of professional engineering in India, he moved to United States to earn PhD Electrical Engineering from Texas A\&M in 2004 following which he relocated to New Zealand (NZ). Nirmal has held various industry, research and academic positions during his career. Currently, he is Associate Dean Postgraduate at the Faculty of Engineering of University of Auckland. 
His research interests span power systems in the context of protective relaying, electricity markets, voltage security, blackouts and resilience. His current focus is towards integration of distributed/renewable energy sources to electricity system with emphasis on protection (IEC 61850, SPS, WAPS), energy markets (block-chain), innovations (Micro-grid, Storage, EV \& PV integration, cyber-resilience, digital twins, machine learning and AI), low-carbon transitions and energy policy. 
He is involved in several leadership positions with IEE and engaged technically with other national and international power engineering organizations like EEA-NZ, CIGRE and CSEE.
\end{IEEEbiography}

\begin{IEEEbiography}[{\includegraphics[width=1in,height=1.25in,clip,keepaspectratio]{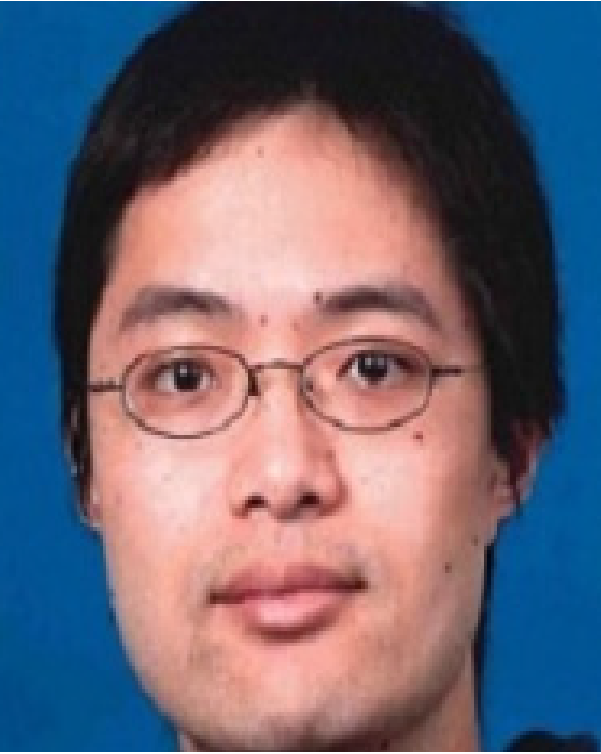}}]{Kevin I-Kai Wang}(Member, IEEE) received the B.E. degree (Hons.) in computer systems engineering and the Ph.D. degree in electrical and electronics engineering from the Department of Electrical and Computer Engineering, The University of Auckland, Auckland, New Zealand, in 2004 and 2009, respectively.

He was a Research Engineer designing commercial home automation systems and traffic sensing systems from 2009 to 2011. He is currently a Senior Lecturer with the Department of Electrical and Computer Engineering at the University of Auckland. His current research interests include wireless sensor network-based ambient intelligence, pervasive healthcare systems, human activity recognition (HAR), behaviour data analytics, and
biocybernetic systems.
\end{IEEEbiography}

\end{document}